\documentclass{aa}

\usepackage{graphicx}
\usepackage[varg]{txfonts}

\newcommand{\gsim}{\hbox{\rlap{\lower.55ex\hbox{$\sim$}} \kern-.3em
\raise.4ex \hbox{$>$}}}
\newcommand{\lsim}{\hbox{\rlap{\lower.55ex\hbox{$\sim$}} \kern-.3em
\raise.4ex \hbox{$<$}}}
\newcommand{\hi}{\textsc{H\,i}}

\begin{document}
   \title{Measuring Diffuse Interstellar Bands with cool stars.}
   \subtitle{An improved line list to model the background stellar spectra}
   \author{A. Monreal-Ibero \inst{1} \and R. Lallement\inst{1}
   \fnmsep\thanks{Based on data products from observations made with ESO Telescopes at the La Silla Paranal Observatory
under programme IDs 66.D-0457(A), 079.C-0131(A), and 383.C-0170(A).}
          }        
   \institute{GEPI, Observatoire de Paris, PSL Research University, CNRS, Universit\'e Paris-Diderot, Sorbonne Paris Cit\'e, Place Jules Janssen,
92195 Meudon, France\\
   \email{[ana.monreal-ibero,rosine.lallement]@obspm.fr}\\
             }             
   \date{Received 20 September 2016; accepted 01 December 2016}

  \abstract
   {Diffuse Stellar Bands (DIBs) are ubiquitous in stellar spectra.
   Traditionally, they have been studied through their extraction from
   hot (early-type) stars, because of their smooth continuum.
   In an era where there are several going-on or planned massive Galactic surveys using  multi-object spectrographs, cool (late-type) stars constitute an
   appealing set of targets. However, from the technical point of view,
   the extraction of DIBs in their spectra is more challenging due to
   the complexity of the continuum.}
   {In this contribution we will provide the community with an
   improved set of stellar lines in the spectral regions associated to
   the strong DIBs at $\lambda$6196.0, 
   $\lambda$6269.8, $\lambda$6283.8, and $\lambda$6379.3.
   These lines will allow for the creation of better stellar synthetic spectra, reproducing the background emission  and a more accurate extraction of  the magnitudes associated with a
   given DIB (e.g. equivalent width, radial velocity).
    }
   {The Sun and Arcturus were used as representative examples of dwarf
   and giant stars, respectively. A high quality spectrum for each of them was modeled
   using TURBOSPECTRUM and the VALD stellar line list.
   The oscillator strength $\log(gf)$ and/or wavelength of specific lines were modified 
   to create  synthetic spectra where  the residuals in both the Sun and Arcturus were minimized.
   }
   {The TURBOSPECTRUM  synthetic spectra based on the improved line lists reproduce the observed spectra for the Sun and Arcturus in the mentioned spectral ranges with greater accuracy. 
   Residuals between the synthetic and observed spectra are always $\lsim$10\%, much better than with previously existing options.
    The  new line list has been tested with some characteristic spectra, from a variety of stars, including both giant and dwarf stars, and under different degrees of extinction.
    As it happened with the Sun and Arcturus residuals in the fits used to extract the DIB information are  smaller when using synthetic spectra made with the updated line lists.
   Tables with the updated parameters are provided to the community. 
   }
   {}

   \keywords{ISM: lines and bands -- ISM: structure  -- Stars: late-type
               }
               
   \maketitle
%
%

\section{Introduction}

Stellar spectra may display some non-stellar weak absorption features of unknown origin associated to one or several clouds of Interstellar Medium (ISM) in their line of sight. 
These are the so-called Diffuse Interstellar Bands (DIBs)  \citep[see][for a review]{her95,sar06}. 
They were already noticed around the early 20's by \citeauthor{heg22} \citep[see][for a revision of the history of the DIBs discovery]{mcc13} and 
their interstellar origin was established in the 30's
\citep{mer34,mer36}.

Today, we know more than 400 of these features \citep[e.g.][]{Galazutdinov00,hob09}. Most of them are seen in the optical, with some additional DIBs clearly identified in the near infrared  \citep[][]{Joblin90,Foing94,Cox14,Hamano16} and a few proposed candidates in the near-UV \citep{Bhatt15}.
Also, DIBs seem omnipresent. Even if the vast majority of DIB research is restricted to our Galaxy, they have been detected in many kind of extragalactic sources, such as the Magellanic Clouds, M31, M33 in the Local Group  \citep{Ehrenfreund02,Welty06,Cordiner08,Cordiner11,Cox07,vanLoon13}, nearby reddened galaxies \citep{Ritchey15}, dusty starburst galaxies \citep{Heckman00,MonrealIbero15}, quasars \citep[e.g.][]{Lawton08}, and supernovae \citep[][]{Sollerman05,Phillips13}.

Still, almost  one century after their discovery, the nature of their carrier(s) (i.e. the agent that originates these features) remains being a mystery \citep[see][and references therein]{Fulara00}.
Among the possible carrier candidates, one can find hydrocarbon chains \cite[e.g.][]{mai04}, polycyclic aromatic hydrocarbons \citep[PAHs, e.g.][]{Salama96,Kokkin08}, and/or fullerenes \citep[][]{igl07,sas01}.
In general, Carbon seems somehow involved. Particularly promising in this regard is the recent confirmation in the laboratory of  C$^+_{60}$ as the carrier of the two DIBs at 9577 and 9632 \AA\, \citep{Campbell15}, confirming an earlier proposal by \citet{Foing94}. 

Not all the DIBs vary in unison. Instead, DIBs are grouped in families. 
Different pairs of DIBs show a range in the degree of correlation when their strengths (as traced by their equivalent widths) are compared, with DIBs in the same family having larger degrees of correlation  \citep[e.g.][]{Cami97,Friedman11,xia12}.
More noteworthy, although the degree of correlation between DIB equivalent width and the column density of molecular hydrogen, $N(H_2)$, is quite variable and depends on the feature under consideration, DIBs present good correlations with  the amount of neutral hydrogen along a given line of sight, the extinction and the interstellar Na\,I\,D and Ca\,H\&K lines \citep[e.g.][]{Herbig93,Friedman11,Lan15,Baron15a}. Thus, irrespective of the actual nature of carrier(s), DIBs can be used as tools to infer properties of the 3D structure of the ISM.

Traditionally, investigations on the nature of the DIBs and their relation with the ISM in general are done using hot (early-type) star spectra since they are brighter and present a spectrum dominated by a smooth, feature-less continuum. However, DIBs seem sensitive to the radiation field of these stars \citep{Vos11,Dahlstrom13,Cordiner13}, and therefore hot stars are not the optimal targets  to prove the typical conditions of the general ISM of our Galaxy.
In fact, \citet{Raimond12} showed how using cooler target stars correlations between DIBs (and reddening) improve, confirming that the radiation field of UV bright stars has a significant influence on the DIB strength. 
Moreover, hot stars are not automatically abundant in a given region of interest.
An alternative strategy would be the use of extra-galactic spectra \citep{Lan15,Baron14}. Although they have in principle  lower quality in terms of signal-to-noise ratio and/or spectral resolution, they are also much more numerous and thus, DIBs can be detected through stacking of similar spectra in terms of extinction and location.

A last option would be using the information carried in cool stars spectra since they are much more abundant and prove less extreme conditions of the ISM in terms of radiation field.
Specifically, thanks to the advent of Multi-Object Spectrographs, we live in an epoch where large spectroscopic surveys offer the possibility of doing so at Galactic scales. Encouraging results in this direction have been presented by most of the on-going surveys as SDSS \citep{Yuan12}, RAVE \citep{Munari08,Kos14} and Gaia-ESO \citep{Puspitarini15} in the optical, as well as SDSS-III APOGEE \citep{Zasowski15,Elyajouri16} in the infrared.
The role of DIBs as tools to gain insight into the Galactic ISM structure will become even more prominent in the forthcoming years now that Gaia satellite provides with a three-dimensional map of the Galaxy, including accurate positions of about one billion stars and hence line of sight directions. Also, with an spectrograph at a resolving power of $\sim$11\,500 \citep{Katz04}, it will provide with accurate measurements of the DIB at $\lambda$8621. Likewise several foreseen highly multiplexed MOSs  (e.g. WEAVE@WHT \citep{Dalton14}, MOONS@VLT \citep{Cirasuolo14}, 4MOST@VISTA \citep{deJong14}) will provide with an immense amount of cool star spectra.  These collections of data will constitute juicy material for DIBs (and ISM, in general) research, and therefore, as preparatory work, it is necessary and timely to revisit and improve the existing methods for extraction of the information associated to the DIBs.

One strategy to  reproduce (and get rid out of) the stellar component  would be the use of synthetic stellar models. With this idea in mind, \citet{Chen13} developed a method to automatically fit the spectra to a combination of a stellar synthetic spectrum, the atmospheric transmission and a given DIB empirical profile. This method was applied later on by  \citet{Puspitarini15} to study the variation of the DIBs as a function of the distance along the LOS as well as to study the DIB-extinction relationship in different regions of the Milky Way. 
Both works pointed out a difficulty with this approach: some of the stellar features were not properly reproduced by the synthetic spectral modeling, thus adding uncertainty to the derivation of the magnitudes associated to a given DIB of interest.
This point has also recently been raised by  \citet{Kohl16} who, after a careful modeling of the stellar emission in the vicinity of the DIB at $\lambda$5780, found no significant DIB absorption in any of their target stars and attributed the differences between modeled and observed spectra  to inaccuracies in the stellar atmospheric modeling rather than to DIB absorption.
The aim of this paper is improving that modeling by revisiting the stellar line list in the spectral regions associated to some of the strongest DIBs. 

Sec. \ref{data} presents the observational data that were used to improve and test the line list. Also it describes our criteria to select  the spectral ranges that we plan to improve.
Sec. \ref{seclinelist} describe our working strategy and provides with a list of modified stellar lines.
Some examples illustrating the improvements for extraction of DIB parameters are included in Sec. \ref{secteststar}.
Our main conclusions are summarized in Sec. \ref{secconclu}.

%
\section{Observational data \label{data}}

We used two sets of data. The first one is made out of two high quality spectra of the Sun and Arcturus and were used to improve the line list. The second one was used to evaluate the quality of the modeling using the new list. Following, details about both sets are presented.

\subsection{The Sun and Arcturus spectra \label{datasunandarc}}

The solar observations are based on fifty solar Fourier Transform Spectrometer (FTS) scans taken by James Brault and Larry
Testerman at Kitt Peak between 1981 and 1984. The spectral resolving power is $\delta\lambda/\lambda\,\sim\,300\,000$.
The signal-to-noise ratio, S/N, is on the order of 3\,000 around 625\,nm.
Details on the spectra can be found in \citet{Kurucz05}.

The spectrum of Arcturus was downloaded from the UVES-POP database
\citep{Bagnulo03}\footnote{http://www.eso.org/sci/observing/tools/uvespop.html.html}.
Its resolution is about 80\,000, being acquired with a 0\farcs{5} slit.

\subsection{The test spectra \label{sectestspectra}}

For this evaluation we intentionally chose a dataset representative of current observing programs with 8-m class telescopes. 
We selected eight  spectra from the ESO data archive to test the modeling of the stellar background emission.
The observing programs they belong to, aim at measuring the metallicity of open cluster members \citep{Santos09,Santos12}.
In order to test a variety of conditions, the spectra correspond to both giants (like Arcturus) and dwarfs (like the Sun) and suffer from extinction in different degree. All of them were obtained with the UVES spectrograph \citep{Dekker00} at the VLT but with a diversity of spectral resolutions. All the spectra cover from 4780 to 6805 \AA. In Table \ref{testspectra} we compile the utilized spectra and their relevant instrumental characteristics.

%
\begin{table*} \caption{Test spectra.} \label{testspectra}
\centering \begin{tabular}{l c c c ccccccccc }     
     
\hline
\hline                      
Star &
Prog. ID &
R=$\lambda/\Delta\lambda$ &
Slit width ($^{\prime\prime}$) &
S/N\tablefootmark{a}  &
Ref. &
$A_v$\tablefootmark{b} &
$E(B-V)$\tablefootmark{c} &
D\tablefootmark{d} (pc) &
Ref.
\\
\hline
NGC\,2682\,Sanders1092    &  66.D-0457   & 87410   & 0\farcs4 & $\sim$80 & S09 & 0.094 & 0.06 & 986 & P10\\
NGC\,2682\,Sanders1048    &  66.D-0457   & 87410   & 0\farcs4 & $\sim$80 & S09 & 0.094 & 0.06 & 986 & P10\\
NGC\,2682\,No164               &  079.C-0131 & 45990   & 0\farcs9 & 100-200  & S09 & 0.094 & 0.06 & 986 & P10\\
NGC\,2682\,No266               &  079.C-0131 & 45990   & 0\farcs9 &100-200   & S09 & 0.094 & 0.06 & 986 & P10\\
IC\,4651\,AMC1109              &  66.D-0457   & 87410   & 0\farcs4 & $\sim$80 & S09 & 0.663 & 0.12 & 888 & K05 \\
IC\,4651\,AMC4226              &  66.D-0457   & 87410   & 0\farcs4 & $\sim$80 & S09 & 0.663 & 0.12 & 888 & K05\\
NGC\,6705\,No1111              &  383.C-0170 & 107200 & 0\farcs3 & 200          & S12 & 2.521 & 0.43 &1877 & K05\\
NGC\,6705\,No1184              &  383.C-0170 & 107200 & 0\farcs3 & 200         & S12  & 2.521& 0.43 &1877 & K05\\
\hline
\end{tabular}
\tablefoot{
S09: \citet{Santos09} ; S12: \citet{Santos12} ; P10: \citet{Pandey10} ; K05: \citet{Kharchenko05}\\
\tablefoottext{a}{Signal-to-noise ratio as provided in the reference quoted in the sixth column.}
\tablefoottext{b}{Total galactic extinction as provided by the NED using the \citet{Schlafly11} recalibration of the \citet{Schlegel98} infrared-based dust map.}
\tablefoottext{c}{Reddening of the cluster as provided in the reference quoted in the last column.}
\tablefoottext{d}{Distance to the cluster  as provided in the reference quoted in the last column.}

}

\end{table*}

%
\begin{table*} \caption{Stellar parameters for Sun and Arcturus.} \label{sunandarcpara}
\centering \begin{tabular}{l c c c ccccccccc }     
     
\hline
\hline                      
Star &
$T_{eff}$\tablefootmark{a} &
$\log g_{spec}$\tablefootmark{a} &
[Fe/H] \tablefootmark{a}  &
[$\alpha$/Fe] &
$\xi_{turb}$ &
Ref. &
Geometry &
R$_{eff}=(\lambda/\Delta\lambda)_{eff}$ \\
&
(K)          &
(cm s$^{-2}$) &
&
&
(km s$^{-1}$)&
&
 \\
\hline
Sun & 5777. & 4.44 & +0.00 & 0.0 & 1.0 & G08 & Plane Parallel & 85\,000\\
Arcturus  & 4247. & 1.54 & -0.52  & 0.2  & 1.3 & J14 &  Spherical &57\,000\\ 
\hline \end{tabular}
\tablefoot{
G08: \citet{Gustafsson08};
J14: \citet{Jofre14};
}
\end{table*}

   \begin{figure*}[!th] \centering
   \label{figoldvsnew}
   \includegraphics[width=0.49\hsize, bb=40 0 790 400, clip=]{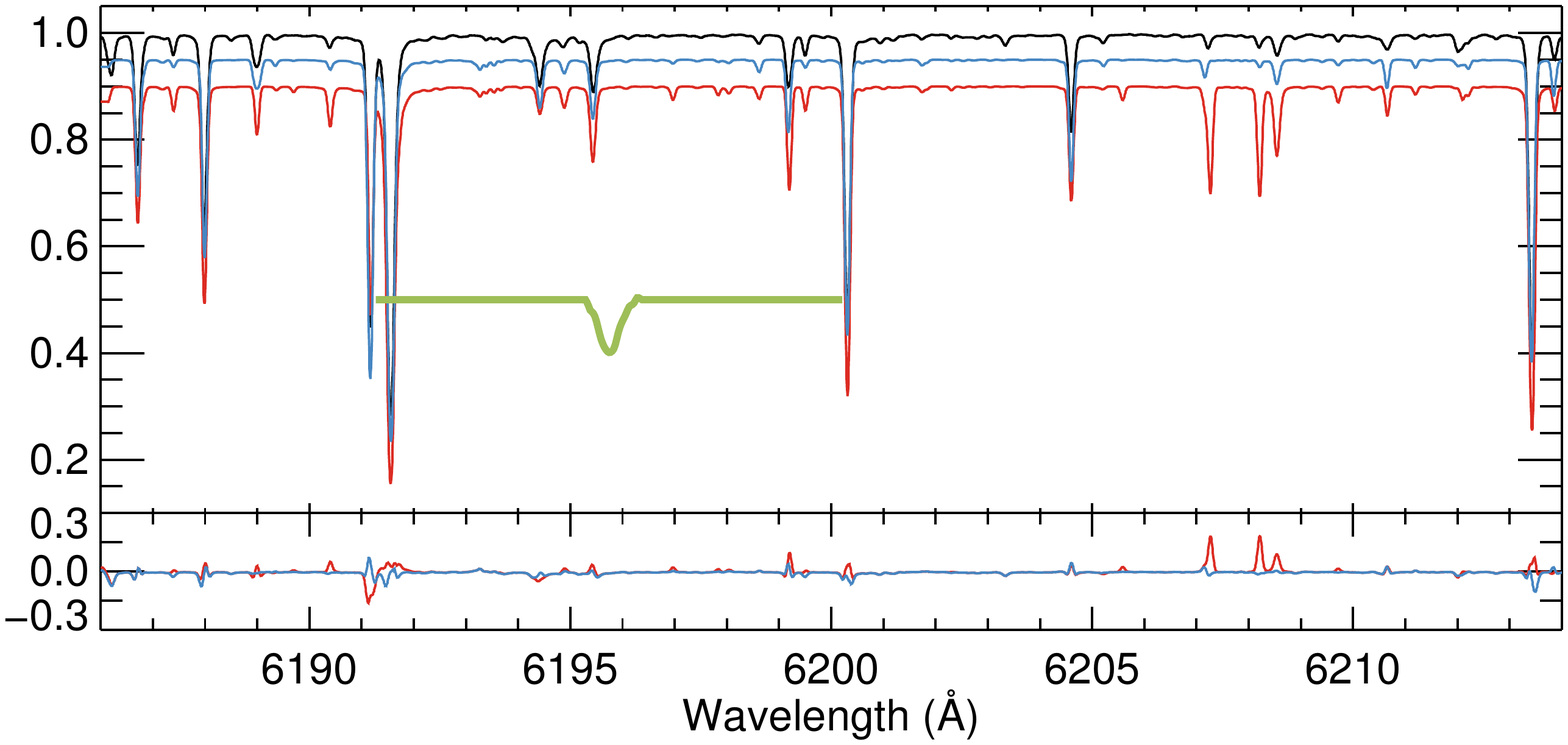}
   \includegraphics[width=0.49\hsize, bb=40 0 790 400, clip=]{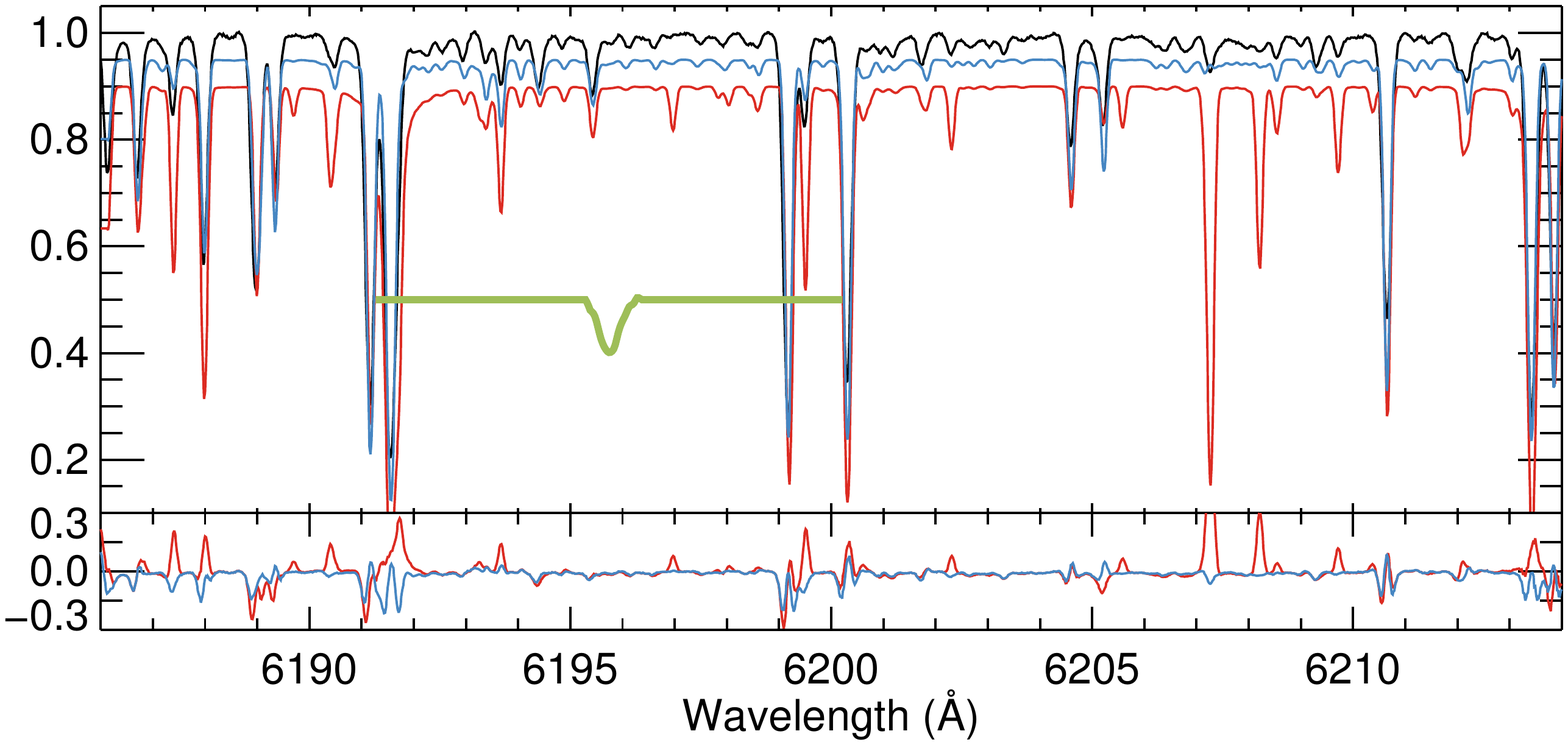}
   \includegraphics[width=0.49\hsize, bb=40 0 790 400, clip=]{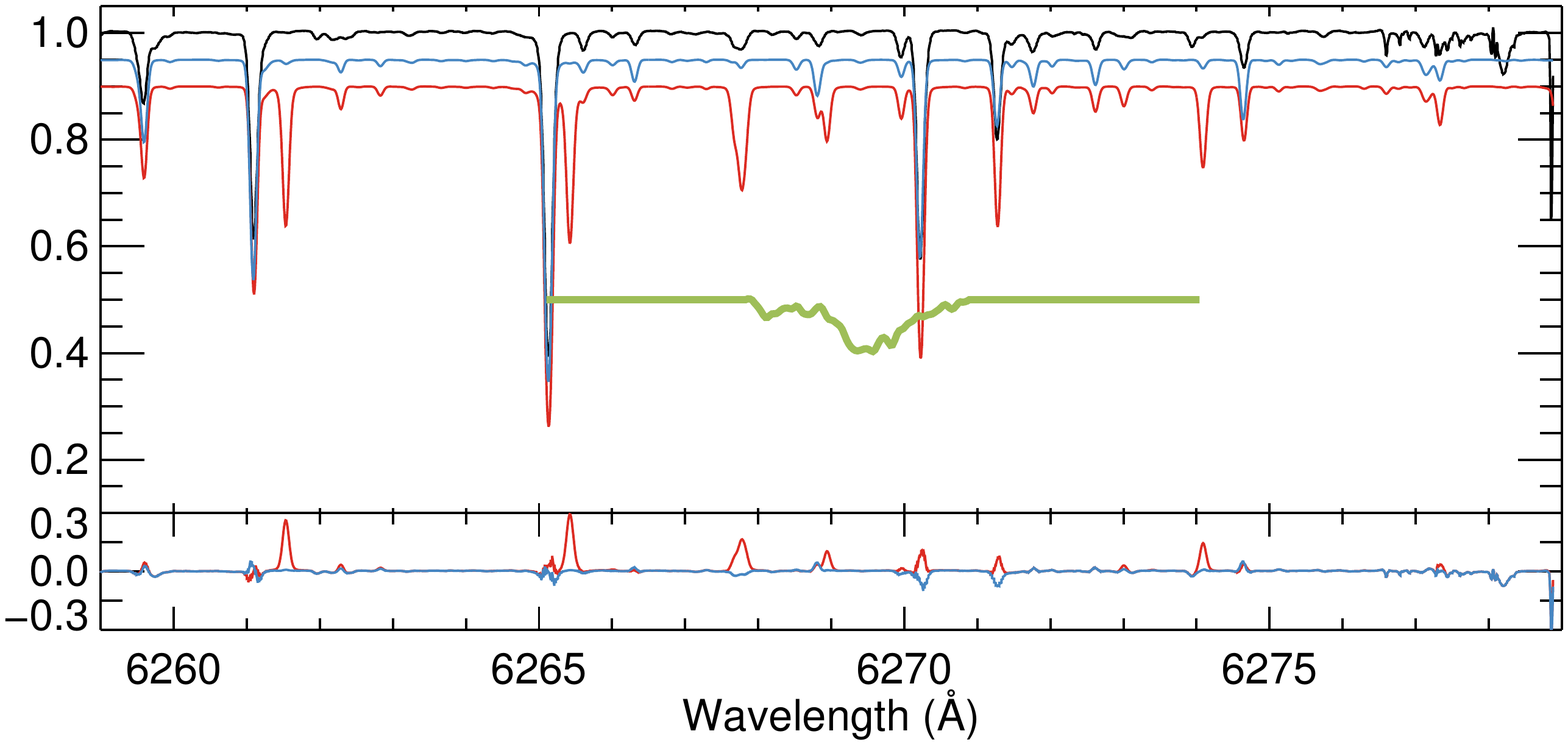}
   \includegraphics[width=0.49\hsize, bb=40 0 790 400, clip=]{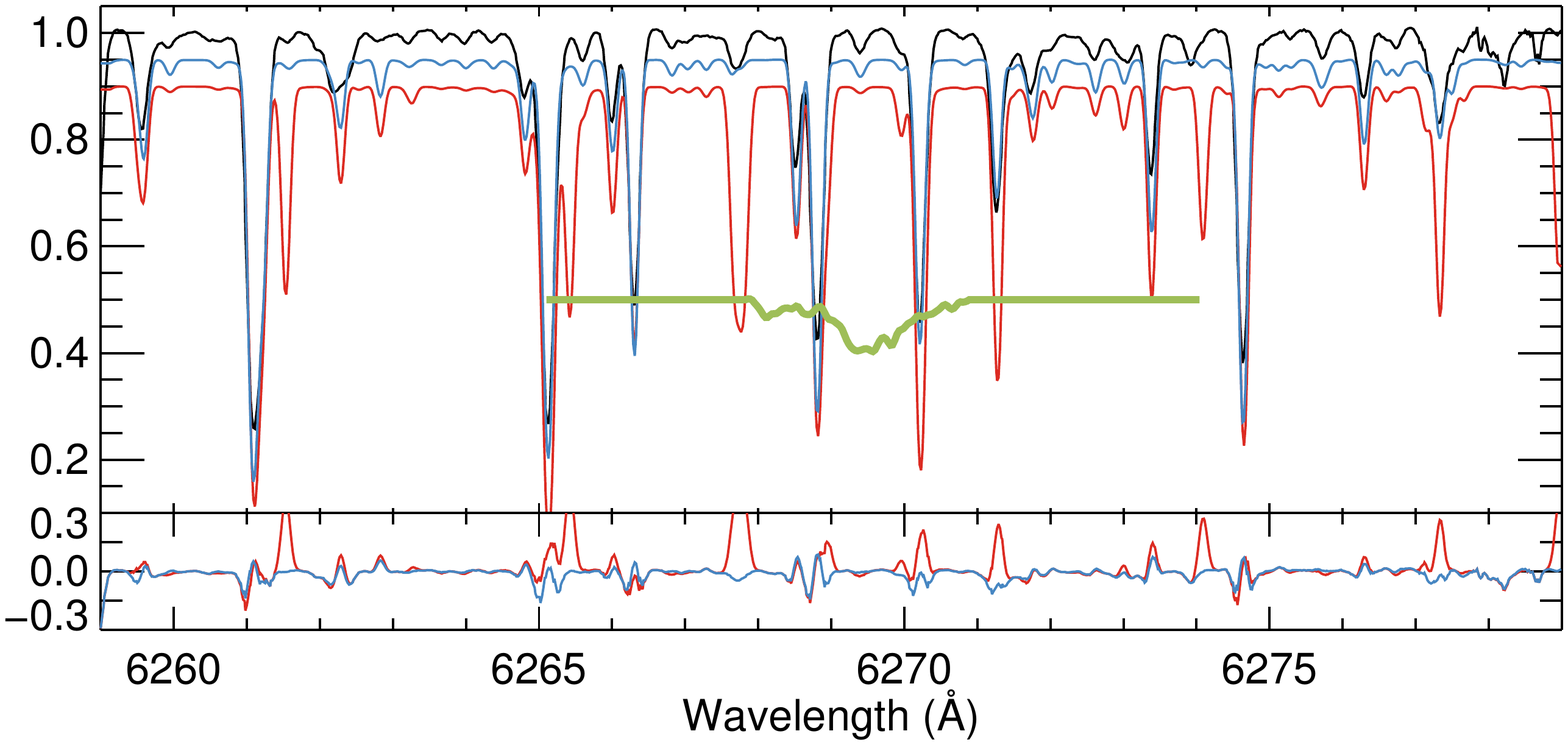}
   \includegraphics[width=0.49\hsize, bb=40 0 790 400, clip=]{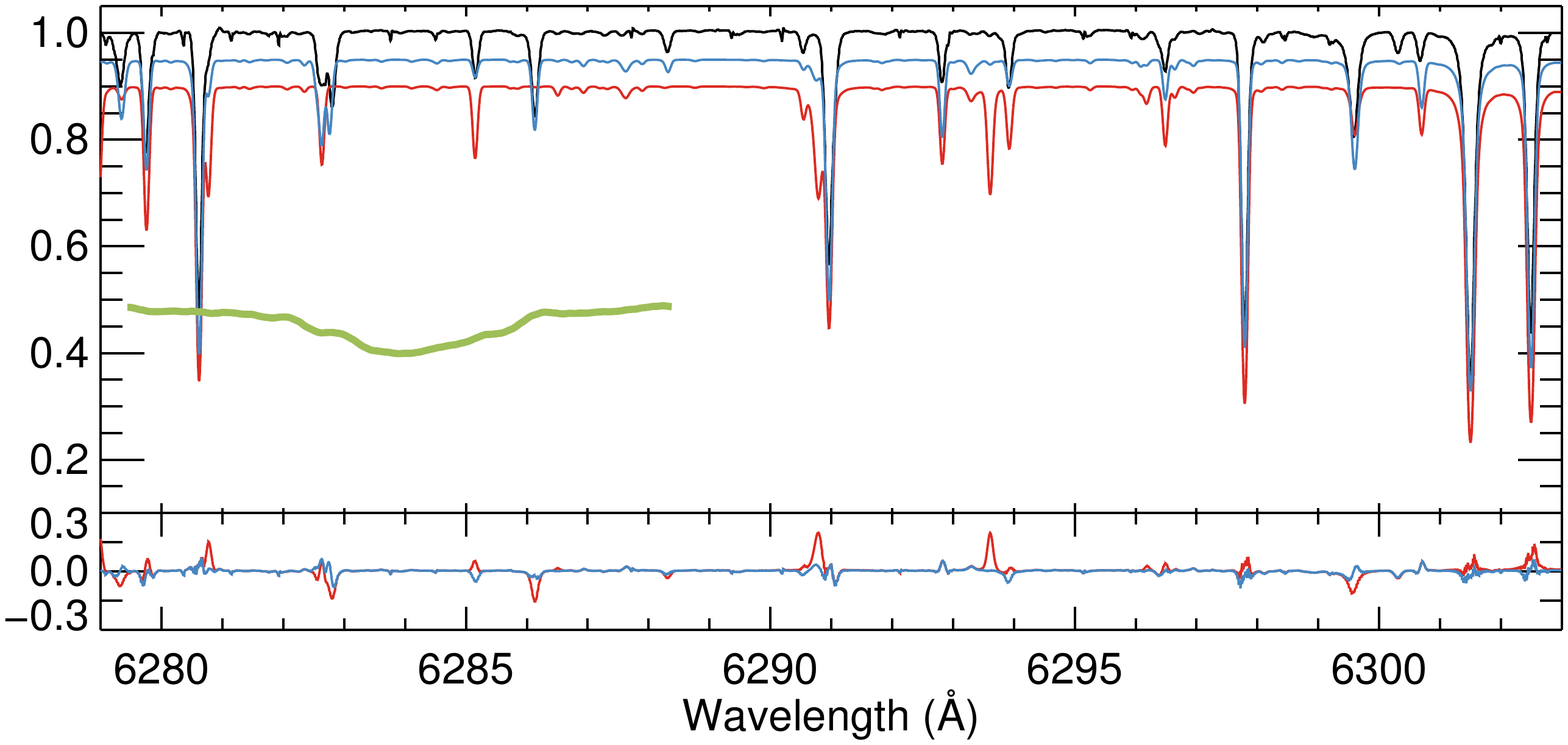}
   \includegraphics[width=0.49\hsize, bb=40 0 790 400, clip=]{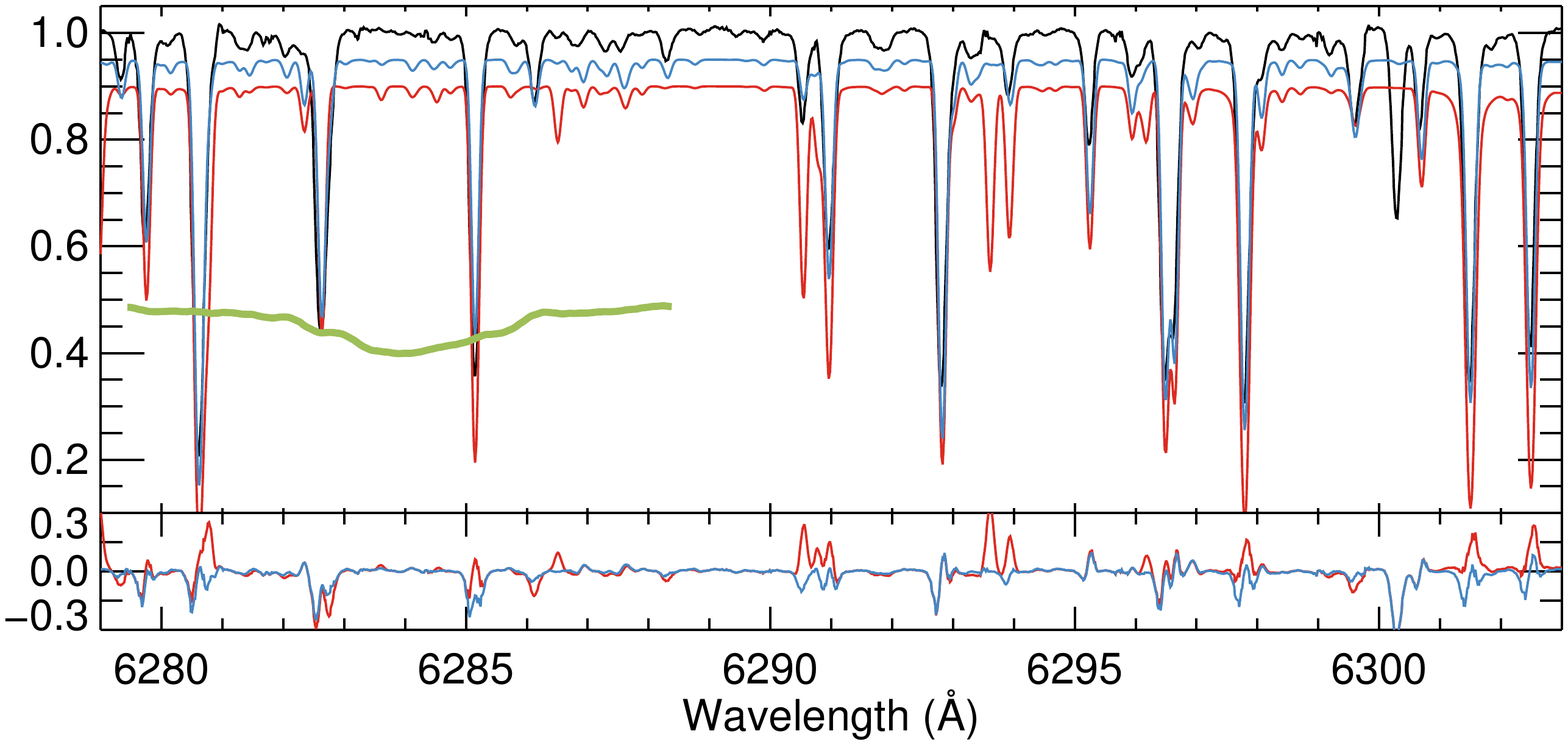}
   \includegraphics[width=0.49\hsize, bb=40 0 790 400, clip=]{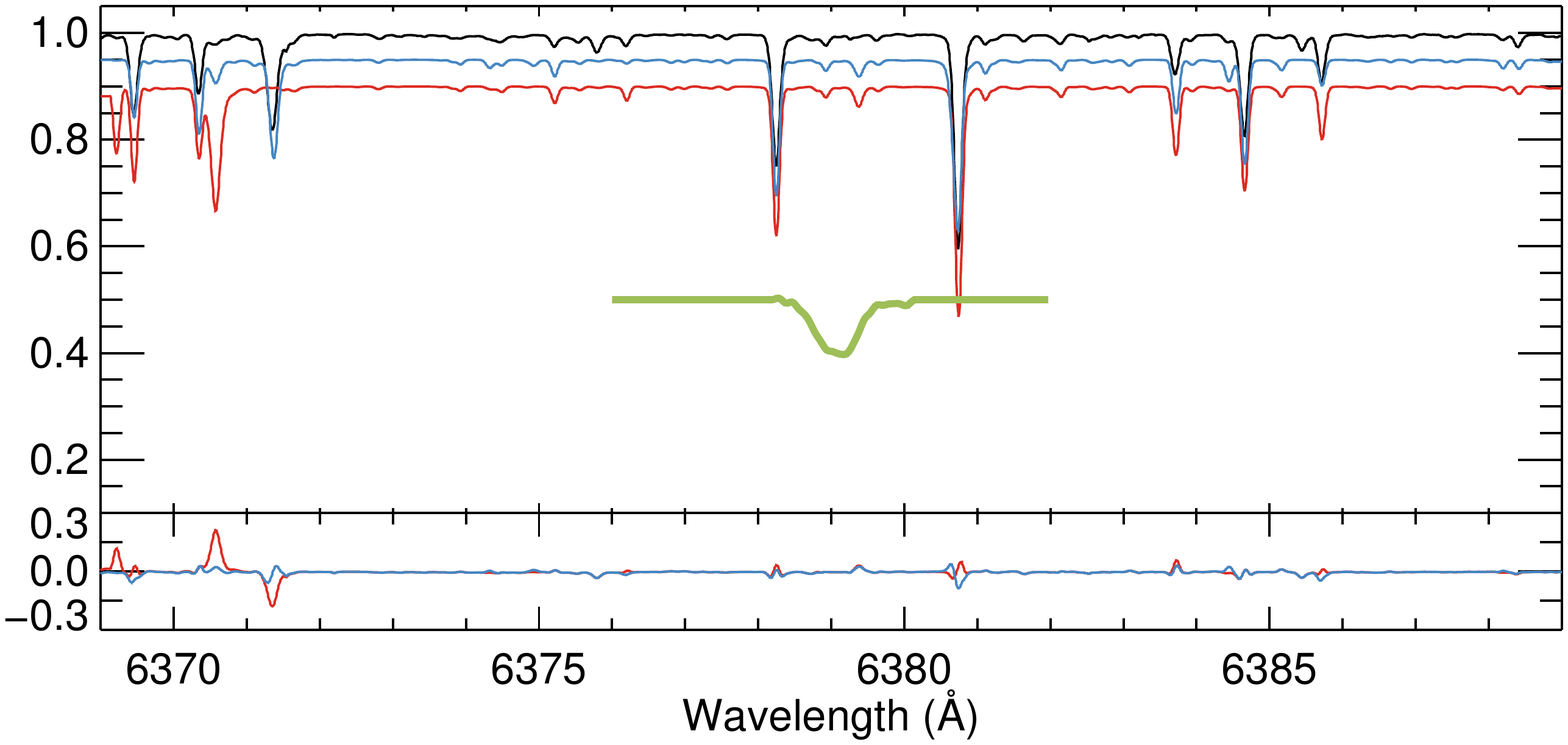}
   \includegraphics[width=0.49\hsize, bb=40 0 790 400, clip=]{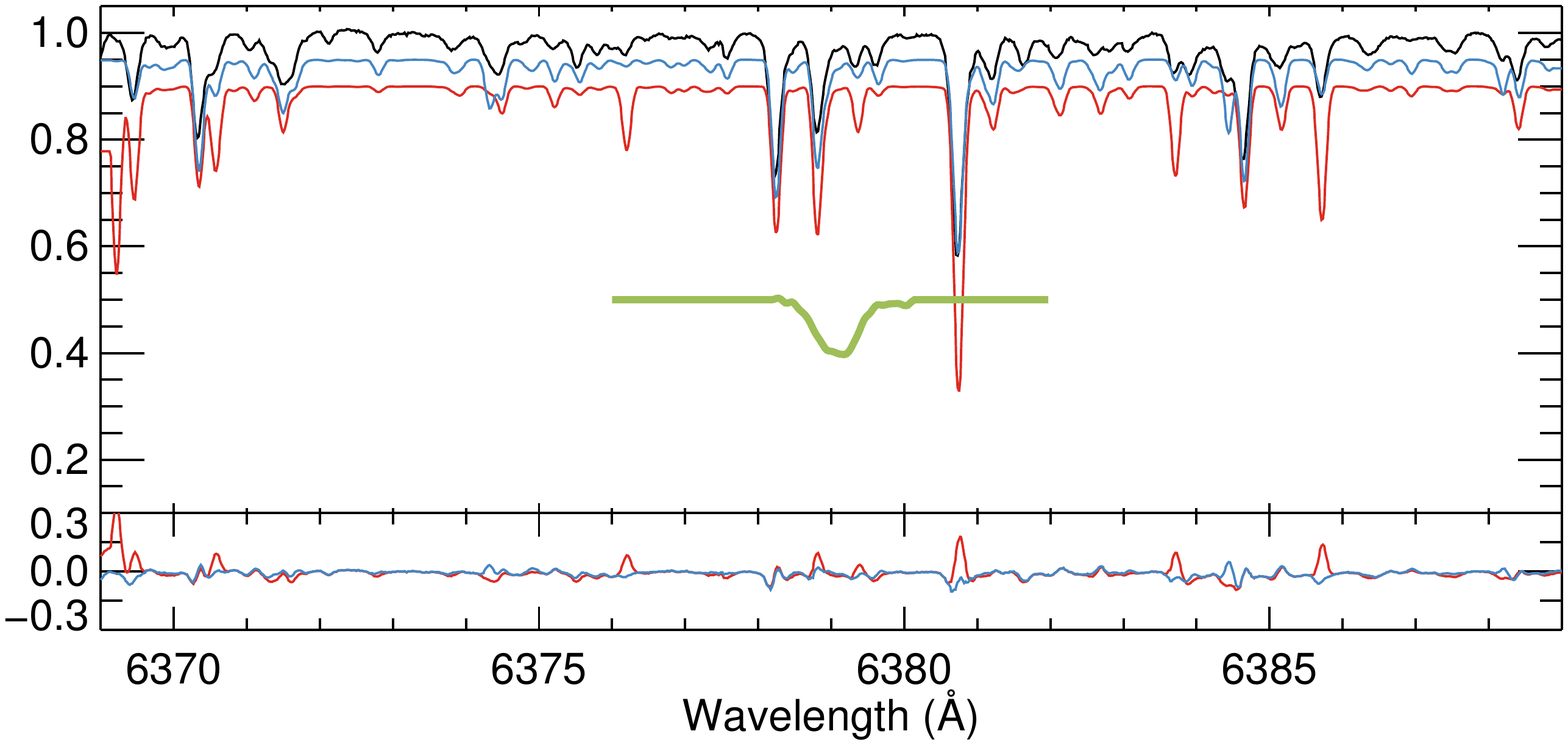}
\caption{Comparison of the modeled spectra of the Sun (\emph{left}) and Arcturus (\emph{right}).
Each pannel contains two graphics.
In the main (\emph{upper}) one, black line indicates the observed spectrum. Red and blue lines have been utilized to show the modeled spectrum using the original and improved line list, respectively. Note that for the shake of clarity these two lines have been offset by $-0.10$ and $-0.05$ in the y-axis.
In the complementary (\emph{lower}) one, the residuals (data - model) using both line lists are shown.
Finally,  the profiles for the DIBs that motivated the corrections presented here and a putative cloud at $v=0$~km~s$^{-1}$ are included for reference.  }
\end{figure*}


\subsection{Studied spectral ranges}

In this contribution we focus on the spectral ranges of a reduced but relevant set of DIBs.
As starting point, we used the list of DIBs presented in Tab. 1 of \citet{Puspitarini13} and 
restricted to those DIBs included in  the spectral range used  by the Gaia-ESO survey to observe bright late-type stars. Specifically, this survey uses for this purpose the UVES spectrograph in its DIC1/580~nm setting, covering from 4760~\AA\ to 6840~\AA. Additionally, we required them to be relatively strong. That means having not only large equivalent width (EW) but also small full width at half maximum (FWHM). Using the results by \citet{Hobbs08} for HD~204827 as reference, we restricted to those DIBs with EW(m\AA)/FWHM(\AA)$>$60.
Also, a good pre-existing model of the DIB profile is needed.
We used the templates  empirically derived  by \citet{Raimond12} and \citet{Puspitarini13} who averaged several FEROS (R$\sim$48000) spectra of early-type (B-A5) stars. Because of their empirical nature, they take into account possible asymmetries in bands and/or blending with neighboring ISM features.
We excluded $\lambda$6445.3 since the model needed further improvements \citep[see Fig. 6. in][]{Puspitarini13}.
In this way, the final list of DIBs of interest includes $\lambda$6196.0, $\lambda$6269.8, $\lambda$6283.8, $\lambda$6379.3, and   $\lambda$6613.6. 
For typical ISM velocities this last feature is strongly blended with a set of stellar lines at $\lambda$6614.4~\AA\, (\ion{Fe}{i}) and 6613.7~\AA\, (\ion{Fe}{i} and \ion{Y}{ii}) particularly difficult to reproduce with the simple strategy described in Sec. \ref{seclinelist}. Therefore, we decide not to consider it at this stage and envision a more refined strategy for this range in the future.
Finally, the fitting procedure needs some continuum towards the blue and red of the DIB under analysis and DIBs trace ISM at a certain velocity (and therefore can appear blue and red-shifted in the spectra). Taken all this into account, the following three spectral ranges were selected for inspection: 6\,186-6\,214~\AA, 
6\,259-6\,303~\AA, 6\,369-6\,389~\AA.

The first spectral range is also needed when analysing the $\lambda\lambda$6203.0,6204.5 blend. 
This is particularly interesting since the relative intensities of the DIBs in the blend do not vary in accord, pointing towards a different origin \citep{Porceddu91}.
Since these two DIBs did not fulfill our criterion of EW(m\AA)/FWHM(\AA) ($\lambda$6204.5 is particularly broad), they will not be tested here. 
However, we note that studies on this blend based on late-type star spectra may equally benefit from the line list provided here.

Also, it is worth to note that the second spectral range is infested with a series of strong absorption telluric features between $\sim$6275~\AA\, and $\sim$6303~\AA\, caused mainly by O$_2$ in the atmophere (but also water vapor). Before tuning the stellar line list, the spectra of the Sun and Arcturus were corrected from this telluric absorption using TAPAS\footnote{http://www.pole-ether.fr/tapas/project?methodName=home\_en} \citep{Bertaux14}.

\begin{table} \caption{Residual statistics for the Sun and Arcturus.} \label{statsunandarcturus}
\scriptsize
\centering \begin{tabular}{llccccccccccc }     
\hline
\hline
  & & \multicolumn{2}{c}{The Sun} & \multicolumn{2}{c}{Arcturus} \\
Range (\AA) &                      
 &
$\mu\pm\sigma$&
Median &
$\mu\pm\sigma$&
Median\\
\hline
6\,186-6\,214 & O & -0.003$\pm$0.021 & -0.004 &  0.001$\pm$0.071 & -0.007\\ 
                      & U & -0.007$\pm$0.012 & -0.005& -0.020$\pm$0.036 & -0.010 \\ 
\hline
6\,259-6\,279 & O &  0.008$\pm$0.034 &  0.003 &  0.013$\pm$0.078 & -0.001\\ 
                      & U & -0.001$\pm$0.017 &  0.002 & -0.013$\pm$0.032 & -0.003\\ 
\hline
6\,279-6\,303 & O &  0.005$\pm$0.029 &  0.004 &  0.001$\pm$0.065 &  0.001\\ 
                      & U &  0.000$\pm$0.014 &  0.003 & -0.016$\pm$0.048 & -0.002\\ 
\hline
6\,369-6\,389 & O & -0.003$\pm$0.024 & -0.003 & -0.007$\pm$0.038 & -0.010\\ 
                      & U & -0.004$\pm$0.010 & -0.003 & -0.011$\pm$0.017 & -0.008\\ 
\hline
\end{tabular}
\tablefoot{
O: Original; U: Updated.\\
}
\end{table}

\begin{table} \caption{List of modified stellar lines  in the $6186-6214$~\AA\ spectral range.} \label{modifiedlines6186_6214}
\centering \begin{tabular}{lcccccccccccc }     
     
\hline
\hline                      
Ion &
$\lambda_{ori}$ &
$\lambda_{new}$ &
$\log(gf)_{ori}$  &
$\log(gf)_{new}$\\
&
(\AA) &
(\AA) &
&
\\
\hline
\hline
\ion{Si}{i}  & 6194.416 & \ldots &-2.076  &-1.800 \\
\ion{Si}{i}  & 6194.884 & \ldots &-2.192 & -2.400 \\
\ion{Si}{i}  & 6195.433 & \ldots & -1.490 & -1.700 \\
\ion{Si}{i}  & 6208.541 & \ldots & -1.467 & -2.000\\
\ion{Ca}{i} & 6204.757 & \ldots & -0.276 & -2.000 \\
\ion{Sc}{i} & 6193.666 & \ldots & -2.760 & -10.000 \\
\ion{Sc}{i} & 6210.658 & 6210.648  & -1.529 &  \ldots \\
\ion{Ti}{i} & 6186.141 & \ldots  & -1.270 & -1.650 \\
\ion{Ti}{i} & 6200.318 & \ldots  & -2.300 & +0.300 \\
\ion{Ti}{i} & 6200.227 & \ldots  & -3.035 & -2.450 \\
\ion{V}{i}  & 6188.961 & 6188.940 & -1.062 & -2.650 \\
\ion{V}{i}  & 6189.364 & 6189.342 & -2.970 & -2.700 \\
\ion{V}{i}  & 6190.495 & \ldots & -2.377 & -2.600 \\
\ion{V}{i}  & 6199.197 &6199.177 & -1.300 & -1.500\\
\ion{V}{i}  & 6207.272 & \ldots & -1.370 & \ldots\\
\ion{V}{i}  & 6213.866 & 6213.846 & -2.050 & -1.850\\
\ion{Fe}{i} & 6187.398$^\ast$ & \ldots & -4.148 & -4.700 \\
\ion{Fe}{i} & 6187.989$^\ast$ & \ldots & -1.720 & -2.600 \\
\ion{Fe}{i} & 6190.398 & \ldots & -1.520 & -2.200\\
\ion{Fe}{i} & 6191.558$^\ast$ & 6191.563 &-1.417& -1.700\\
\ion{Fe}{i} & 6196.968 & \ldots & -2.233 & -2.800\\
\ion{Fe}{i} & 6198.042 & \ldots & -2.049 & -2.400\\
\ion{Fe}{i} & 6199.506$^\ast$ &\ldots &-4.430 & -4.975\\
\ion{Fe}{i} & 6200.312& 6200.302 &-2.437& -2.900\\
\ion{Fe}{i} & 6202.305&\ldots &-5.191  & -5.650\\
\ion{Fe}{i} & 6205.585&\ldots &-2.219 & -3.500\\
\ion{Fe}{i} & 6207.230&\ldots &-1.969 &  \ldots\\
\ion{Fe}{i} & 6208.211&\ldots & -1.139 & -2.500\\
\ion{Fe}{i} & 6209.714 &\ldots &3.249  & -3.800 \\
\ion{Fe}{i} & 6212.013 &\ldots &-4.758 & -3.300\\
\ion{Fe}{i} & 6212.099 &\ldots &-2.913 &-3.400\\
\ion{Fe}{i} & 6213.429 &6213.419 &-2.482 &-2.900\\
\ion{Fe}{ii} & \ldots & 6187.996 & \ldots & -4.800 \\
\ion{Fe}{ii} & \ldots & 6199.190 & \ldots & -3.941 \\
\ion{Co}{i} & 6188.996& \ldots & -2.450& -2.500 \\
\ion{Co}{i} & 6197.833& \ldots & -0.510& -1.200 \\
\ion{Ni}{i} & 6191.178 &6191.168 & -2.939 & -2.300  \\
\ion{Ni}{i} & 6204.600 & \ldots & -1.100 & -1.060 \\
\ion{Y}{i} & 6191.718$^\ast$ & \ldots & -0.680 & \ldots \\
\ion{Nd}{ii} & 6210.680 & 6210.660 & -1.540 &-1.540 \\
\hline
\end{tabular}
\tablefoot{
$^\ast$: Lines susceptible to finer tuning (see discussion in Sec. \ref{secteststar}).\\
}
\end{table}

\begin{table} \caption{Modified stellar lines  in the $6259-6303$~\AA\ spectral range.} \label{modifiedlines6259_6303}
\centering \begin{tabular}{lcccccccccccc }     
     
\hline
\hline                      
Ion &
$\lambda_{ori}$ &
$\lambda_{new}$ &
$\log(gf)_{ori}$  &
$\log(gf)_{new}$\\
&
(\AA) &
(\AA) &
&
\\
\hline
\hline
\ion{Si}{i}  & 6279.343 & \ldots & -2.434  & -1.700\\
\ion{Si}{i}  &6290.792 & \ldots &-1.074 & -2.800\\
\ion{Si}{i}  &6299.599 & \ldots &-1.658 & -1.200\\
\ion{Sc}{i} & 6276.295 & \ldots & -2.605 &-2.720 \\
\ion{Sc}{ii} & 6279.753 & \ldots & -1.252 & -1.430 \\
\ion{Ti}{i} & 6261.099 & 6261.089 & -0.530 & -0.470 \\
\ion{Ti}{i} & 6266.010 & \ldots & -1.950& -2.150 \\
\ion{Ti}{i} & 6268.525 & \ldots & -2.260& -2.200 \\
\ion{Ti}{i} & 6273.388 & \ldots & -4.008& -4.180 \\
\ion{Ti}{i} & 6293.004 & \ldots & -3.100& -3.000\\
\ion{Ti}{i} & 6295.248 & \ldots & -4.242& -4.290\\
\ion{Ti}{i} & 6296.646 & \ldots & -3.582& -3.650\\
\ion{V}{i} & 6266.307 & \ldots & -2.290&-2.090 \\
\ion{V}{i} & 6268.798 & \ldots & -2.128&-2.000 \\
\ion{V}{i} & 6274.649 & \ldots & -1.670&-1.620 \\
\ion{V}{i} & 6285.150 & \ldots & -1.510&-2.200 \\
\ion{V}{i} & 6296.487 & \ldots & -1.590&-1.790 \\
\ion{V}{ii} & 6261.087 & \ldots & -2.389&-2.189 \\
\ion{Fe}{i} & 6261.534 & \ldots & -1.004 & -2.800 \\
\ion{Fe}{i} & 6265.132 & 6265.127 & -2.550 & -2.750 \\
\ion{Fe}{i} &6265.422 & 6265.423 & -0.975 & -3.000 \\
\ion{Fe}{i} &6267.676 & \ldots & -2.759 & -5.000 \\
\ion{Fe}{i} &6267.766 & \ldots & -1.363 & -2.500 \\
\ion{Fe}{i} &6267.825 & \ldots & -2.376 & -6.000 \\
\ion{Fe}{i} &6268.942 & 6268.932 & -1.527 & -2.900 \\
\ion{Fe}{i} &6270.224 & 6270.214 & -2.464 & -2.900 \\
\ion{Fe}{i} &6271.278 & 6271.268 & -2.703 & -3.150 \\
\ion{Fe}{i} &6274.089& \ldots & -1.325 & -2.400 \\
\ion{Fe}{i} &6277.334 & \ldots & -4.001 &-4.300 \\
\ion{Fe}{i} &6277.530 & 6277.529 & -4.651 &-4.520 \\
\ion{Fe}{i} & 6278.966 & \dots & -1.139 &-2.900 \\
\ion{Fe}{i} & 6280.770 & \dots & -1.659 &-2.300 \\
\ion{Fe}{i} & \dots & 6282.558  & \ldots & -1.800 \\
\ion{Fe}{i} & \dots & 6282.760  & \ldots & -1.400 \\
\ion{Fe}{i} & 6286.133 & 6286.130  &-3.086 & -3.200 \\
\ion{Fe}{i} & 6286.509 & \ldots  & -3.447 & -4.330\\
\ion{Fe}{i} & 6288.323 & \ldots & -3.845 & -2.900\\
\ion{Fe}{i} & 6290.543 & \ldots & -4.330 & -4.875\\
\ion{Fe}{i} & 6293.611 & \ldots & -1.156 &-2.700\\
\ion{Fe}{i} & 6293.924 & \ldots & -1.717 &-2.150\\
\ion{Fe}{i} & 6296.180 & \ldots & -2.094 &-2.600\\
\ion{Fe}{i} & 6297.792 & \ldots & -2.740 &-3.000\\
\ion{Fe}{i} & 6301.500& \ldots & -0.718 &-1.200\\
\ion{Fe}{i} & 6302.494& \ldots & -0.973  &-1.300\\
\ion{Fe}{ii} & 6269.959& \ldots & -4.500  &-4.800\\
\ion{Fe}{ii} & \ldots  &6286.130  & \ldots  &-2.800\\
\ion{Fe}{ii} & \ldots  &6301.500  & \ldots  &+0.400\\
\ion{Fe}{i} & 6380.743 & \dots & -1.376 & -1.750 \\
\ion{Co}{i} & 6262.829 &  \ldots & -2.644 & -2.800\\
\ion{Co}{i} & 6273.004 &  \ldots & -1.035 & -1.400\\
\ion{Ni}{i} & 6259.595 &  \ldots & -1.237 & -1.300\\
\ion{La}{ii} & 6262.290 & \dots & -1.220  & -1.500 \\
\hline
\end{tabular}
\end{table}

\begin{table} \caption{Modified stellar lines  in the $6369-6389$~\AA\ spectral range.} \label{modifiedlines6369_6389}
\centering \begin{tabular}{lcccccccccccc }     
     
\hline
\hline                      
Ion &
$\lambda_{ori}$ &
$\lambda_{new}$ &
$\log(gf)_{ori}$  &
$\log(gf)_{new}$\\
&
(\AA) &
(\AA) &
&
\\
\hline
\hline
\ion{Si}{i}  & 6370.574 & \ldots & -0.947  & -1.900\\
\ion{Si}{i}  & 6380.689 & \ldots & -2.733 & -1.400\\
\ion{Si}{ii}  & 6371.371 &  \ldots & -0.040 & -0.150\\
\ion{Ca}{i} & 6374.930 & \ldots & -0.525 &-1.500\\
\ion{Sc}{i} & 6378.807 & \ldots & -2.420 & -2.625 \\
\ion{Ti}{i} & 6371.496 & \ldots & -1.940 & -1.900 \\
\ion{Ti}{i} & 6374.321 & \ldots & -3.359 & -0.700 \\
\ion{V}{i} & 6379.364 & \ldots & -0.995 & -1.995 \\
\ion{V}{i} & 6384.445 & \ldots & -0.804 & +0.800 \\
\ion{Fe}{i} & 6369.217 & \ldots & -2.344 & -4.300 \\
\ion{Fe}{i} & 6376.201 & \dots & -2.928 & -3.440 \\
\ion{Fe}{i} & 6380.743 & \dots & -1.376 & -1.750 \\
\ion{Fe}{i} & 6383.708 &  \ldots & -2.644 & -3.100\\
\ion{Fe}{i} & 6385.718 & \dots & -1.910 & -2.200\\
\ion{Fe}{i} & 6388.405 & \dots & -4.476  & -4.270 \\
\ion{Fe}{ii} & 6369.459 & \dots & -4.231  &-4.450 \\
\ion{Ni}{i} & 6370.346 & \dots & -1.940  & -1.890 \\
\ion{Ni}{i} & 6378.247 & \dots & -0.830 & -0.900\\
\ion{Sr}{i} & 6388.199 & \ldots & -1.070 & +0.000\\
\hline
\end{tabular}
\end{table}

\section{Towards an optimized line list  \label{seclinelist}}

The stellar line list was improved by comparing synthetic spectra with the observed spectra for the Sun and Arcturus described in Sec. \ref{datasunandarc}.
For that, we utilized the radiative transfer code TURBOSPECTRUM \citep[][]{Alvarez98,Plez12} that requires a model with the stellar atmosphere structure. For the Sun, we utilized one of the native models from the grid of MARCS models for late-type stars \citep{Gustafsson08} while for Arcturus, we interpolated using a code provided by T. Masseron\footnote{http://marcs.astro.uu.se/software.php}.
Additionally, basic stellar parameters as well as the geometry of the atmosphere needed to be provided. Utilized parameters for both stars are listed in Tab. \ref{sunandarcpara}.
The last input for TURBOSPECTRUM is a list with the physical parameters of the lines that we intend to tune. We used as starting point those provided by the Vienna Atomic Line Database (VALD-3)\footnote{http://vald.astro.uu.se/} \citep{Ryabchikova15}. 
With this, two initial synthetic spectra were created and degraded to an effective resolution that takes into account the broadening due to rotation and turbulence of the stars. Utilized values are showed in the last column of Tab.  \ref{sunandarcpara}. 

The synthetic spectra using TURBOSPECTRUM and the original VALD-3 line list
are shown in Fig. \ref{figoldvsnew} in red.
A quick inspection of this Figure is enough to state that several spectral features  are not properly modeled with residuals with absolute values   $\gsim10$\% .
In most of those cases, equivalent width of the stellar line is overestimated. Some examples have already been noticed  \citep[see e.g. a residual at $\sim$6609.5~\AA\, in Fig. 2 of ][]{Puspitarini15}.

We modified the VALD-3 line list with the esprit of keeping the tuning as simple as possible. Thus, we allowed ourselves to modify only the oscillator strength ($\log(gf)$) and, exceptionally, the wavelength of the line. In those few cases where no line responsible for a given feature was identified, we created a new line by replicating a nearby Fe\,I or Fe\,II line at the wavelength of the feature that we intend to reproduced. This is reasonable strategy, since iron lines are ubiquitous and iron is good proxy for the star metallicity. Finally, in those cases where a different $\log(gf)$ was required to minimize the residuals for the Sun and for Arcturus, we adopted an intermediate value for $\log(gf)$ as a compromise.  This is a pragmatic approach valid for our goals (i.e.  get rid out of the stellar spectra, without necessarily having a deep understanding of the stellar physics).
Synthetic spectra using 
TURBOSPECTRUM and our modified line list are shown in blue in Fig. \ref{figoldvsnew}. Even if there are still some low-level residuals, typically $\lsim$10\%, these synthetic spectra clearly beat those created with the TURBOSPECTRUM code together with the original VALD-3 line list.

Table \ref{statsunandarcturus}, that contains the means, standard deviations and medians of the residuals in each subplot of Fig. \ref{figoldvsnew}, offers a more quantitative way of comparing all the three options. By looking at the standard deviations of the residuals, it is clear that TURBOSPECTRUM models using the updated line list are much better than those using the original one. Since the residuals that matter most are those within the relatively narrow wavelength ranges that include only each DIB itself plus some limited continuum to either side, we also estimated the \emph{local} standard deviation, in 2 \AA-wide bins, for both the Sun and Arcturus. As it happened with the global standard deviations, these  were comparable or smaller  when using the updated line list than when using the original ones.

The updated line lists are the main outcome of this contribution and they are offered to the community in Tables \ref{modifiedlines6186_6214} to \ref{modifiedlines6369_6389}. In the following section, we will show some examples of its applicability for DIB extraction.

\section{DIB extraction using the new line list  \label{secteststar}}

\begin{table} \caption{Stellar parameters  for the test stars.} \label{stellarpara}
\scriptsize
\centering \begin{tabular}{l c c c ccccccccc }     
\hline
\hline                      
Star &
$T_{eff}$\tablefootmark{a} &
$\log g_{spec}$\tablefootmark{a} &
$\xi_{turb}$\tablefootmark{a} &
[Fe/H] \tablefootmark{a}  &
Ref.\\
&
(K)          &
(cm s$^{-2}$) &
(km s$^{-1}$) & 
&
 \\
\hline
NGC~2682 San1092    &6074.   &4.39   &1.35 &+0.04   &  S09\\
NGC~2682 San1048    &5915.    &4.48   & 0.96 &+0.07  &   S09\\
NGC~2682 No164    &4812.   &2.73    & 1.57 &+0.03   &   S09\\
NGC~2682 No266    &4862.   &2.76    & 1.59 &+0.01   &   S09\\
IC~4651 AMC1109    &6075.   &4.54    & 1.14 &+0.15   &   S09\\
IC~4651 AMC4226     &5862.   &4.31    & 0.89 &+0.13   &   S09\\
NGC~6705 No1111    &5039.   &2.85   & 2.18  & +0.14  &    S12\\
NGC~6705 No1184    &4518.   &2.09    & 1.92  &$-$0.01   &   S12\\
\hline \end{tabular}
\tablefoot{
S09: \citet{Santos09} ; S12: \citet{Santos12}\\
\tablefoottext{a}{Using the line-list of \citet{Sousa08}}.
}
\end{table}

\begin{table*} \caption{DIB measurements for the test stars.} \label{resultstestars}
\scriptsize
\centering \begin{tabular}{llccccccccccccccccccccc }     
\hline
\hline
  &  &
  \multicolumn{3}{c}{6196.0} &
  \multicolumn{3}{c}{6269.8} &
  \multicolumn{3}{c}{6283.8} &
  \multicolumn{3}{c}{6379.3}  \\
Star &     &                 
EW&
$v$&
$\chi^2$  &
EW$\pm$e(EW)&
$v$&
$\chi^2$ &
EW&
$v$&
$\chi^2$ &
EW &
$v$&
$\chi^2$   \\
  &  &
 (m\AA) & (km s$^{-1}$) & &
 (m\AA) &  (km s$^{-1}$) &&
 (m\AA) &  (km s$^{-1}$) &&
  (m\AA) &  (km s$^{-1}$)\\
\hline
NGC2682 San1092   & O &   $<10$  &  \ldots & 0.81 & $<28$  & \ldots & 1.00 &  $182\pm18$  & $-9\pm5$ & 1.61 & $<16$ & \ldots & 1.19\\ 
                                  & U &    $<9$  &  \ldots & 0.79 & $<21$ & \ldots& 0.66 &  $120\pm15$ & $+19\pm7$ & 1.35 & $<11$ & \ldots & 0.50\\ 
\hline
NGC2682 San1048   & O &   $<20$  &  \ldots  & 2.78 &  $<72$   & \ldots &  6.19 &  $<194$ & \ldots & 4.73 & $<27$ & \ldots & 2.82\\ 
                                 & U &  $<20$  &  \ldots  & 2.72 &  $<69$   & \ldots  &  5.67 &  $<180$ & \ldots & 4.23 & $<25$ & \ldots & 2.34\\ 
\hline
NGC2682 No164   & O &   $23\pm17$   &  $-4\pm2$  & 0.96 & $<47$  & \ldots & 2.72 & $243\pm93$ & $+13\pm7$  &3.96 & $<14$ & \ldots & 0.90\\ 
                              & U &   $14\pm11$   & $-4\pm3$  & 0.82 & $<22$ & \ldots & 0.65 & $125\pm73$ & $+12\pm8$ &1.53  & $<10$ & \ldots & 0.49 \\ 
\hline
NGC2682 No266   & O &22$\pm$16 &$-4\pm2$ & 0.81 &  $<45$ & \ldots  & 2.51 & $251\pm79$ & $+9\pm6$ & 3.57 & $<14$ & \ldots & 0.88\\ 
                              & U &  14$\pm$10 & $-5\pm3$  & 0.70 & $24\pm20$ &  $-19\pm4$ & 0.60 & $158\pm84$ & $+9\pm6$ & 1.82 &  $<10$ & \ldots & 0.48\\ 
\hline
IC 4651 AMC1109  & O &  16$\pm$2  &  $-15\pm2$ & 0.53 &  $<30$ & \ldots & 1.19 & $265\pm38$ & $+9\pm3$ & 1.63 &   $19\pm3$ & $-15\pm2$ & 0.33\\ 
                              & U &  17$\pm$2   &  $-15\pm2$ & 0.37 & $31\pm18$ & $-18\pm2$ & 0.41 & $189\pm9$ & $+1\pm5$ & 0.70 & $19\pm2$ & $-16\pm2$ & 0.32\\ 
\hline
IC 4651 AMC4226  & O & 12$\pm$3  &  $-17\pm3$ & 0.67 &  $<31$ & \ldots & 1.19 & $75\pm15$ & $-60\pm16$  &2.75 &  $13\pm3$ & $-16\pm3$ & 0.42\\ 
                              & U &  12$\pm$3 & $-16\pm3$  & 0.55 &  $23\pm11$& $-26\pm3$ & 0.51 &  $203\pm9$ & $-28\pm4$  & 1.09 &   $13\pm3$ & $-18\pm3$ & 0.40\\ 
\hline
NGC6705 No1111 & O &  29$\pm$13 &  $-24\pm2$ & 1.90 & $46\pm37$ & $-33\pm4$ & 4.17 & $804\pm51$ & $-45\pm2$ & 4.45 & $79\pm9$ & $-18\pm1$ & 1.99\\ 
                             & U & 36$\pm$10  &  $-22\pm1$ & 1.94 &  $77\pm18$ & $-28\pm2$ & 1.60 &  $736\pm74$ & $ -45\pm2$  &  2.44 & $87\pm7$ & $-19\pm1$ & 1.69\\ 
\hline
NGC6705 No1184  & O &  26$\pm$8  &  $-24\pm2$ & 2.47 &  $78\pm6$ & $-37\pm13$  & 4.25 & $810\pm67$ & $-46\pm2$  &4.75  & $83\pm15$ & $-19\pm1$ & 1.49\\ 
                                & U &  34$\pm$12 & $-23\pm2$  & 2.66 &  $115\pm44$  & $-43\pm2$ & 2.23 & $712\pm111$  & $-44\pm2$ & 3.14 & $101\pm20$ & $-18\pm1$ & 1.58 \\ 
\hline

\end{tabular}
\tablefoot{
O: Original; U: Updated. Reported errors for the DIB velocities include only the formal 1-sigma statistical errors associated to the fit. See main text for a discussion about additional sources of uncertainty.
}
\end{table*}

\begin{figure*}[!th] \centering
   \textsf{\small Dwarf stars\hspace{8cm} Giant stars}\\
   \includegraphics[width=0.49\hsize,  angle=180, bb=5 468 780 572, clip=]{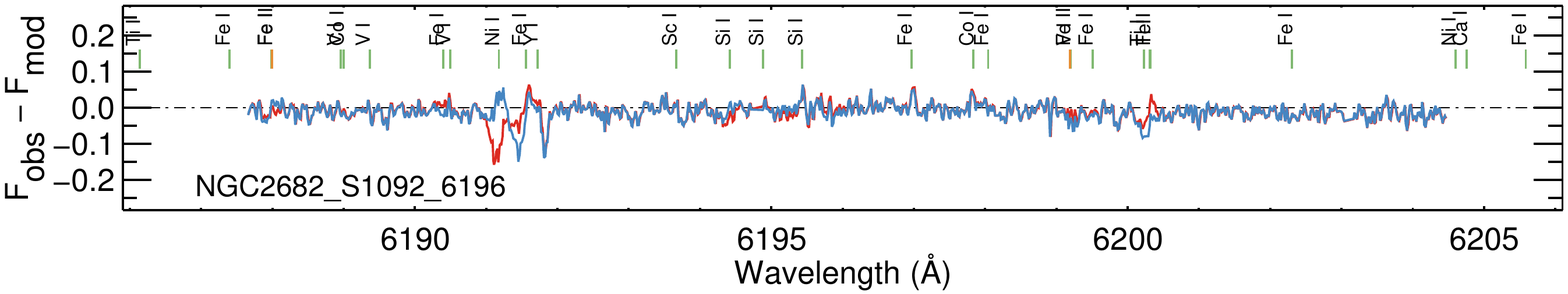}
   \includegraphics[width=0.49\hsize,  angle=180, bb=5 468 780 572, clip=]{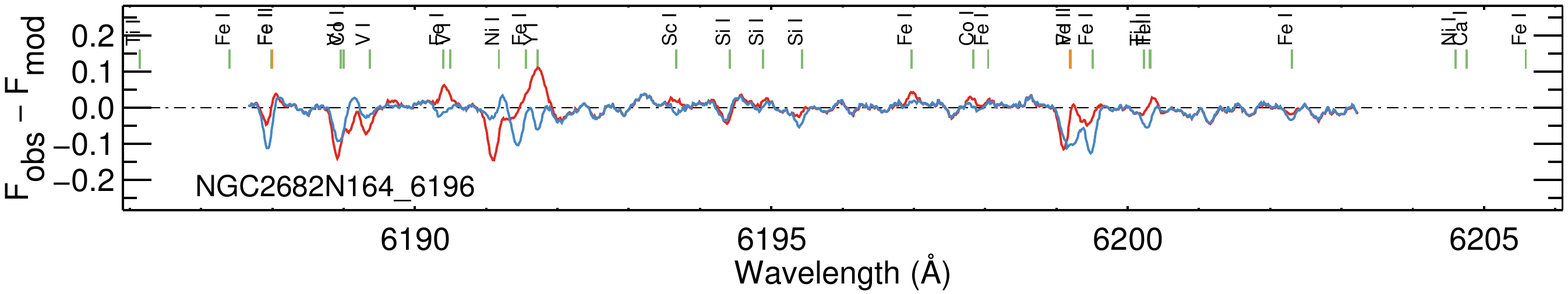}\\
   \includegraphics[width=0.49\hsize,  angle=180, bb=5 468 780 572, clip=]{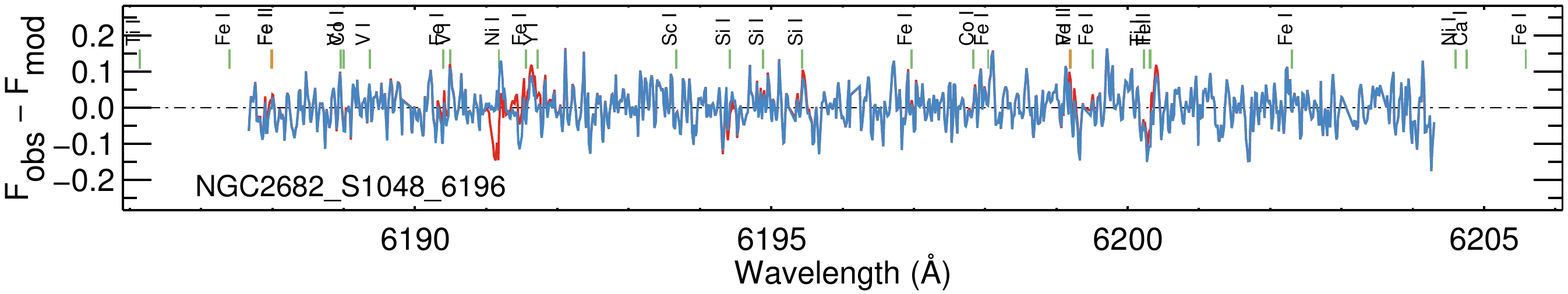}
   \includegraphics[width=0.49\hsize,  angle=180, bb=5 468 780 572, clip=]{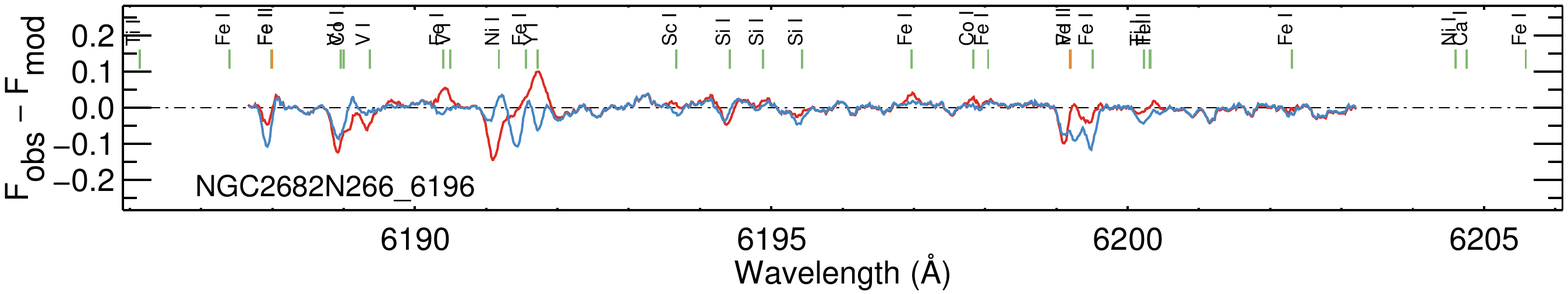}\\
   \includegraphics[width=0.49\hsize,  angle=180, bb=5 468 780 572, clip=]{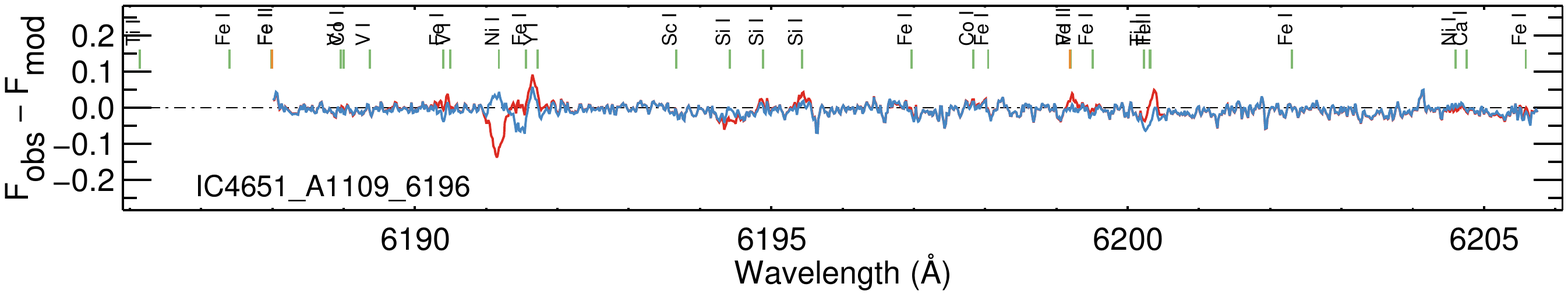}
   \includegraphics[width=0.49\hsize,  angle=180, bb=5 468 780 572, clip=]{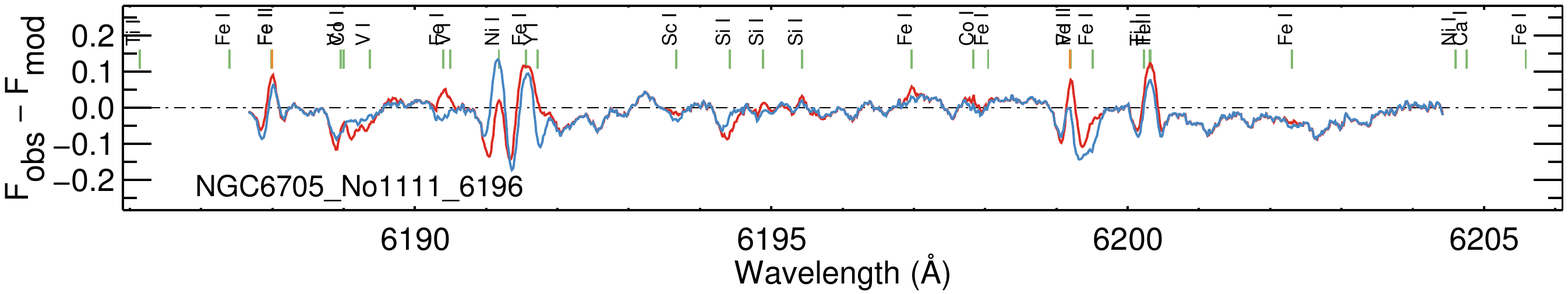}\\
   \includegraphics[width=0.49\hsize,  angle=180, bb=5 468 780 612, clip=]{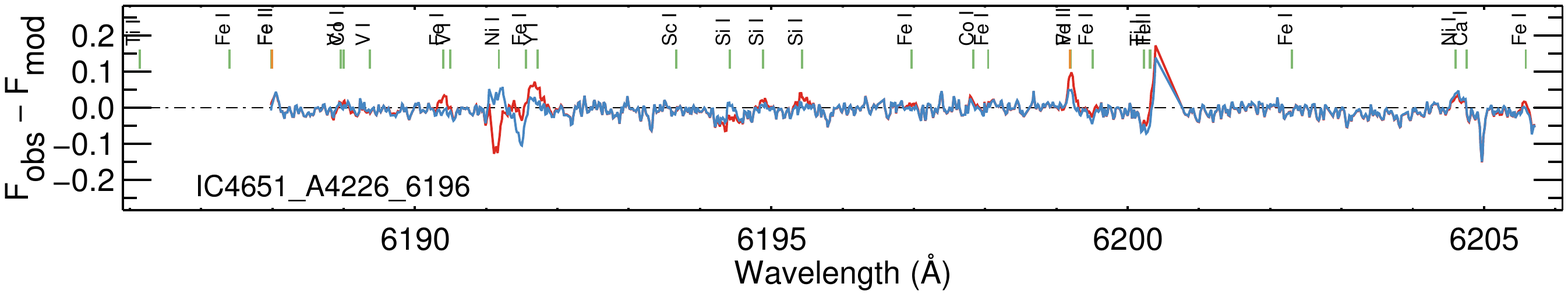}
   \includegraphics[width=0.49\hsize,  angle=180, bb=5 468 780 612, clip=]{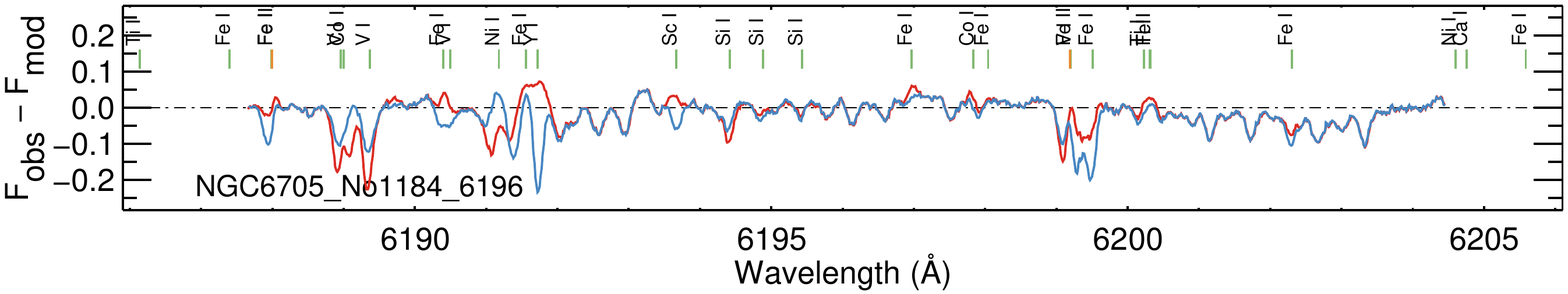}
 \caption{Comparison of the fit residuals for the DIB at $\lambda$6196.0~\AA. Dwarf test stars are on the left column, while giant stars are on the right. We used red lines for the synthetic spectra using the original line list and blue lines for those with the improved one, as in Fig. \ref{figoldvsnew}.}
\label{figresi6196}
 \end{figure*}

\begin{figure*}[!th] \centering
   \textsf{\small Dwarf stars\hspace{8cm} Giant stars}\\
   \includegraphics[width=0.49\hsize,  angle=180, bb=5 468 780 572, clip=]{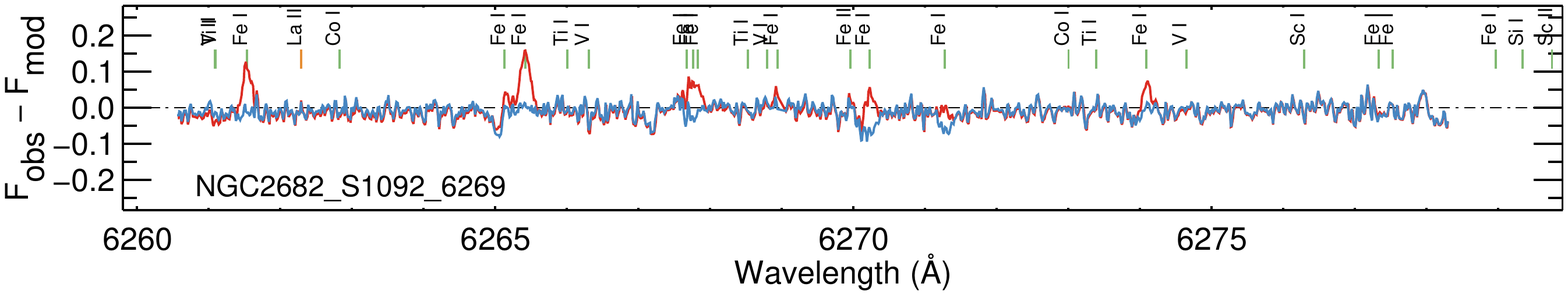}
   \includegraphics[width=0.49\hsize,  angle=180, bb=5 468 780 572, clip=]{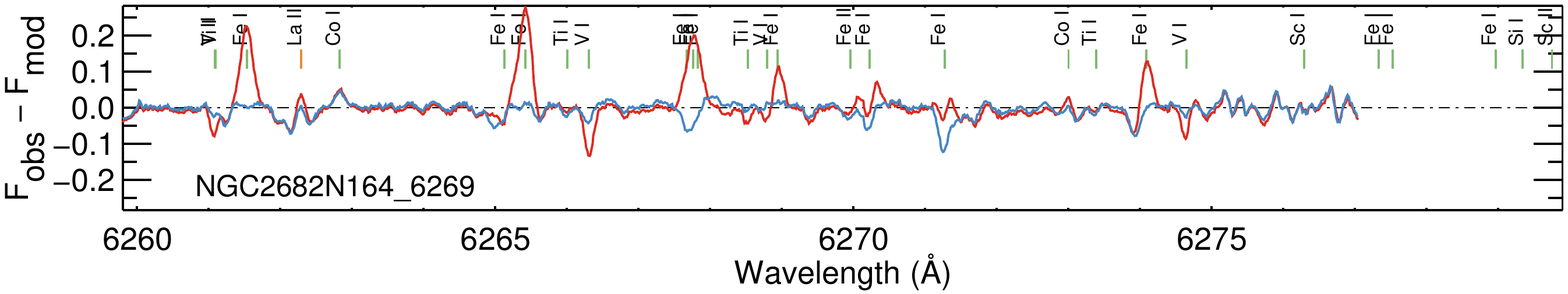}\\
   \includegraphics[width=0.49\hsize,  angle=180, bb=5 468 780 572, clip=]{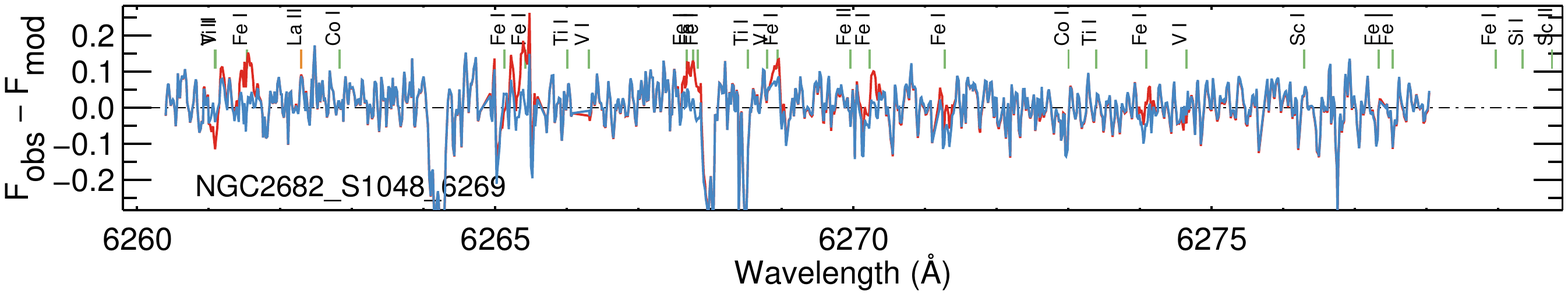}
   \includegraphics[width=0.49\hsize,  angle=180, bb=5 468 780 572, clip=]{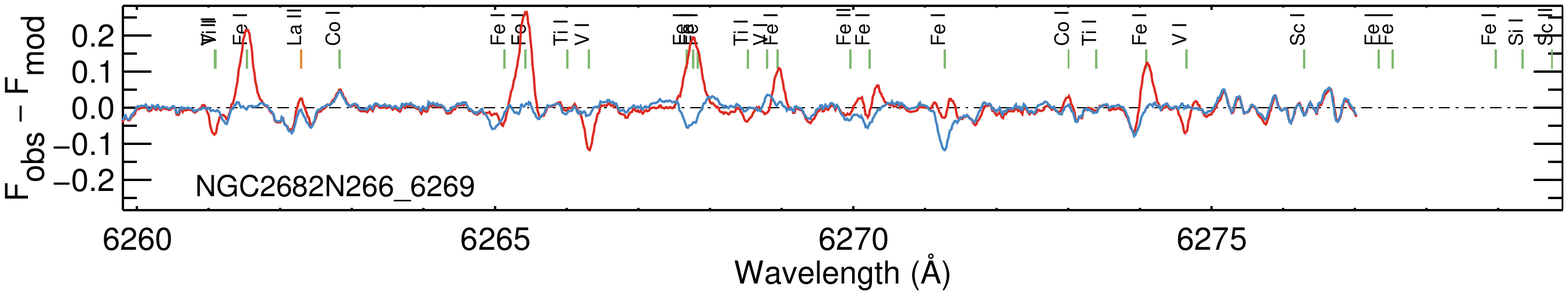}\\
   \includegraphics[width=0.49\hsize,  angle=180, bb=5 468 780 572, clip=]{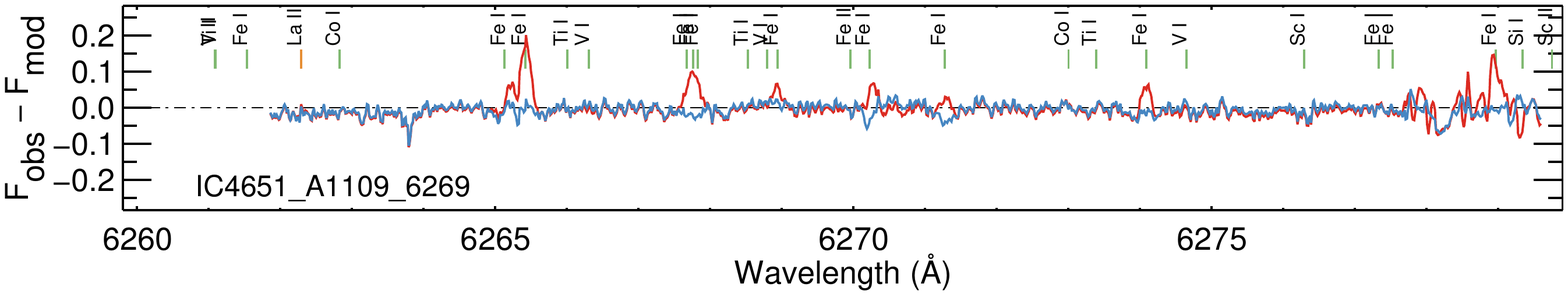}
   \includegraphics[width=0.49\hsize,  angle=180, bb=5 468 780 572, clip=]{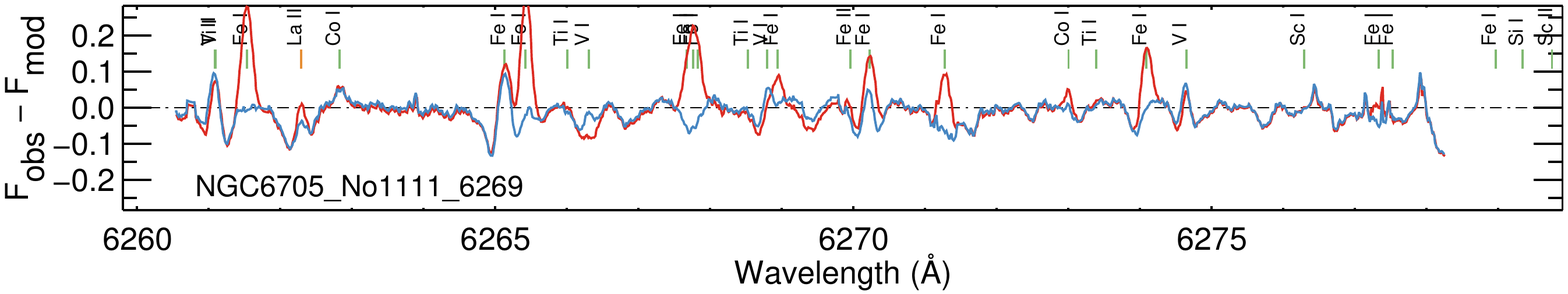}\\
   \includegraphics[width=0.49\hsize,  angle=180, bb=5 468 780 612, clip=]{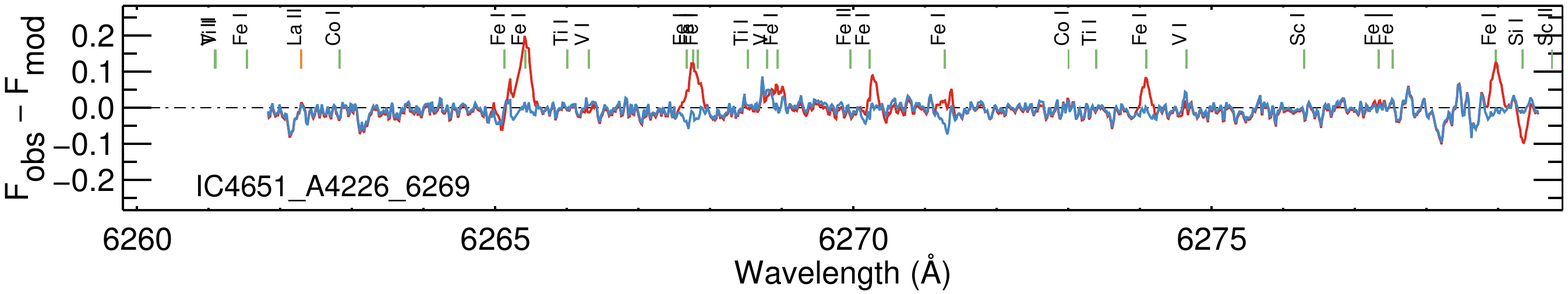}
   \includegraphics[width=0.49\hsize,  angle=180, bb=5 468 780 612, clip=]{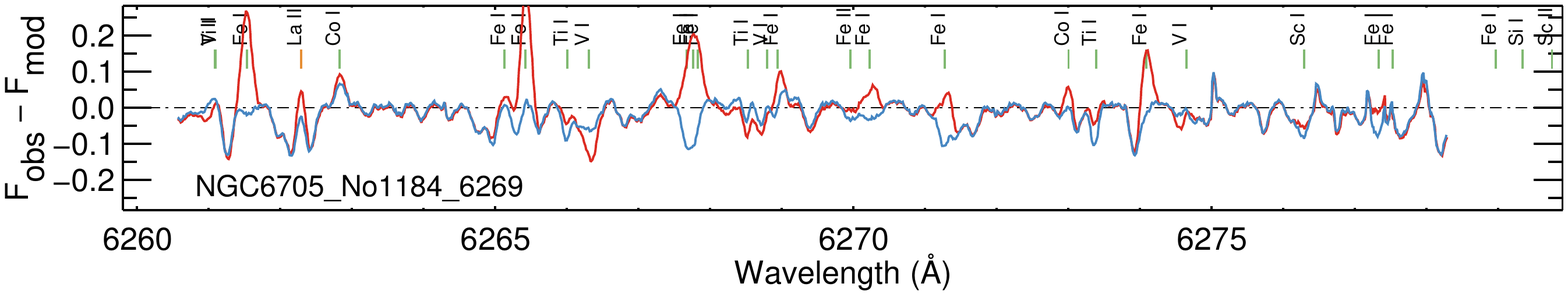}
 \caption{Same as for Fig. \ref{figresi6196}  but for the DIB at $\lambda$6269.8~\AA.}
\label{figresi6269}
 \end{figure*}

\begin{figure*}[!th] \centering
   \textsf{\small Dwarf stars\hspace{8cm} Giant stars}\\
   \includegraphics[width=0.49\hsize,  angle=180, bb=5 468 780 572, clip=]{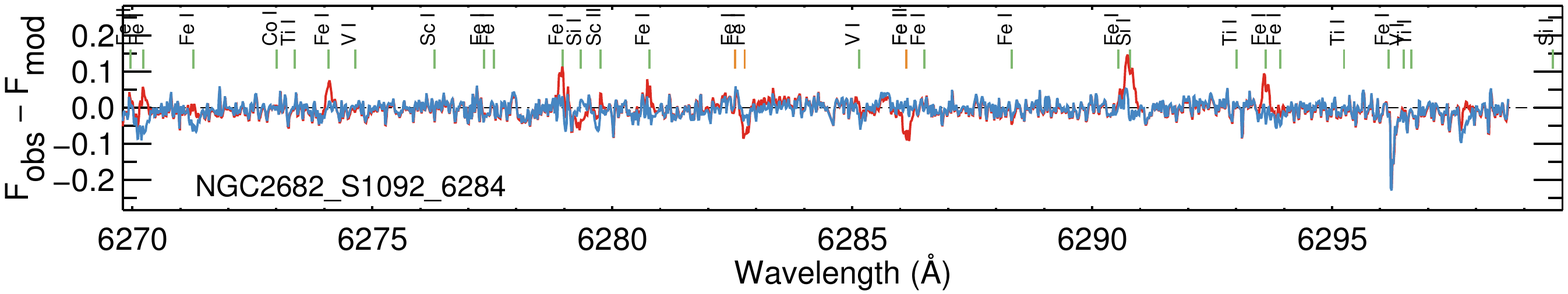}
   \includegraphics[width=0.49\hsize,  angle=180, bb=5 468 780 572, clip=]{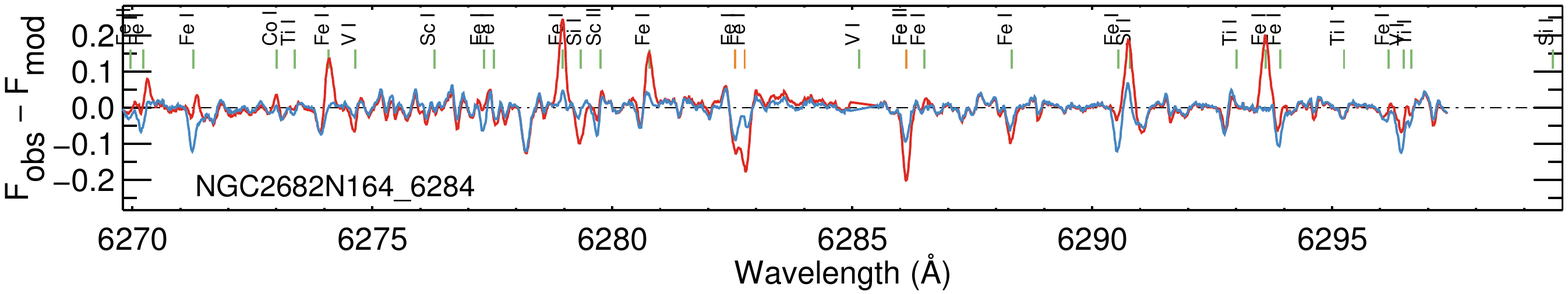}\\
   \includegraphics[width=0.49\hsize,  angle=180, bb=5 468 780 572, clip=]{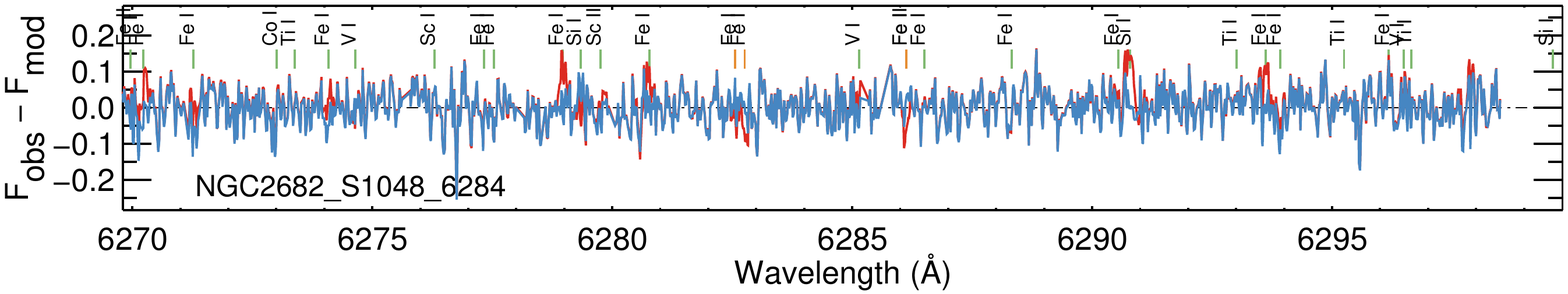}
   \includegraphics[width=0.49\hsize,  angle=180, bb=5 468 780 572, clip=]{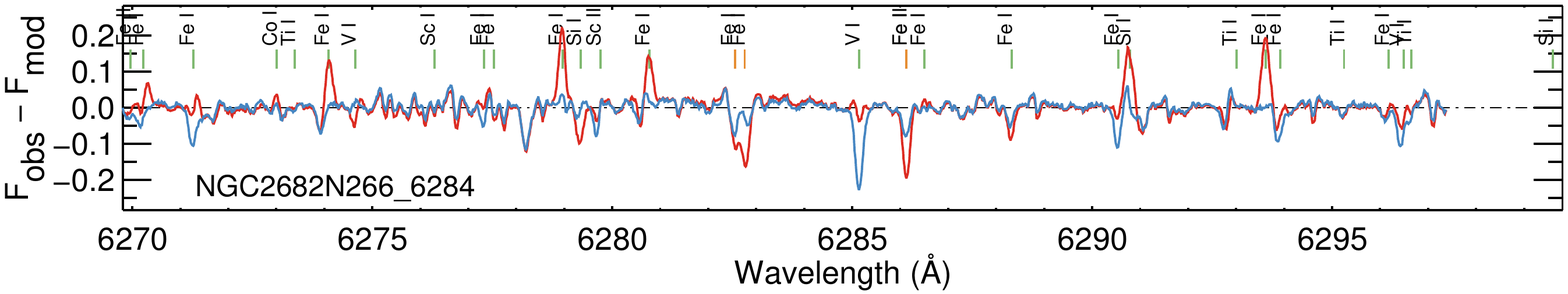}\\
   \includegraphics[width=0.49\hsize,  angle=180, bb=5 468 780 572, clip=]{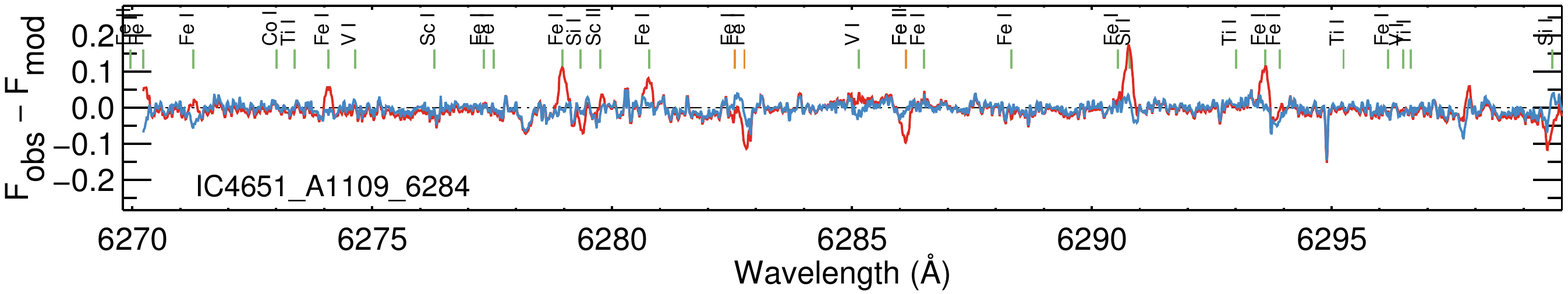}
   \includegraphics[width=0.49\hsize,  angle=180, bb=5 468 780 572, clip=]{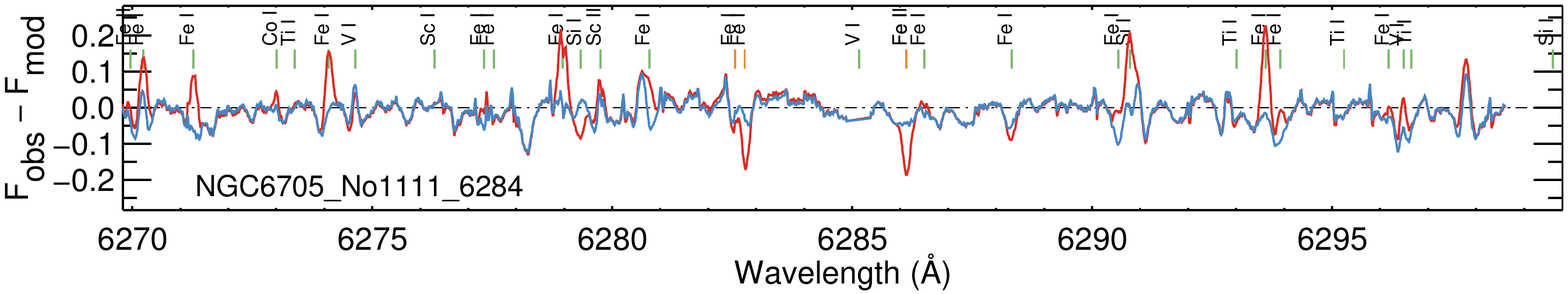}\\
   \includegraphics[width=0.49\hsize,  angle=180, bb=5 468 780 612, clip=]{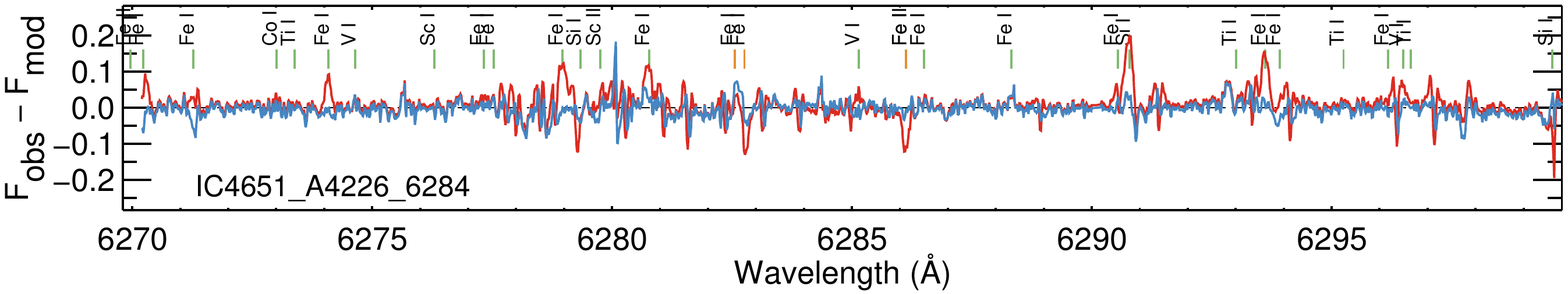}
   \includegraphics[width=0.49\hsize,  angle=180, bb=5 468 780 612, clip=]{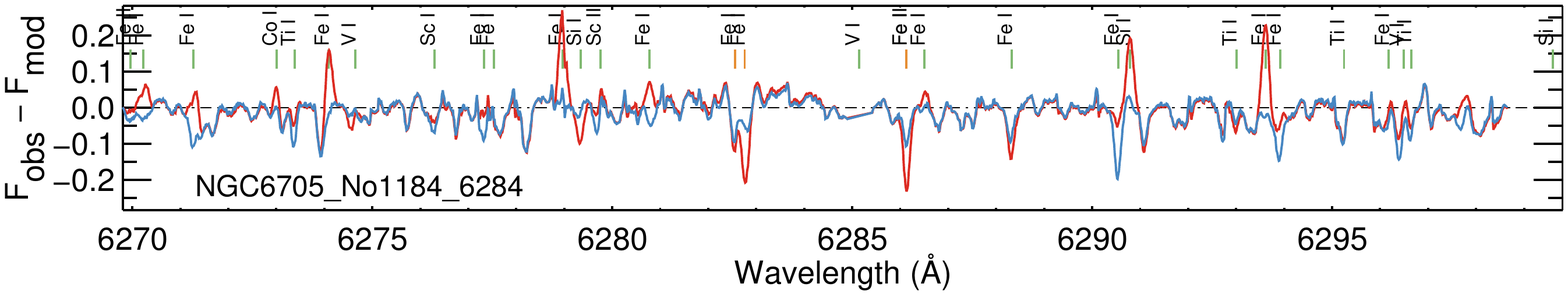}
 \caption{Same  as for Fig. \ref{figresi6196}  but for the DIB at $\lambda$6283.8~\AA. }
\label{figresi6284}
 \end{figure*}

\begin{figure*}[!th] \centering
   \textsf{\small Dwarf stars\hspace{8cm} Giant stars}\\
   \includegraphics[width=0.49\hsize,  angle=180, bb=5 468 780 572, clip=]{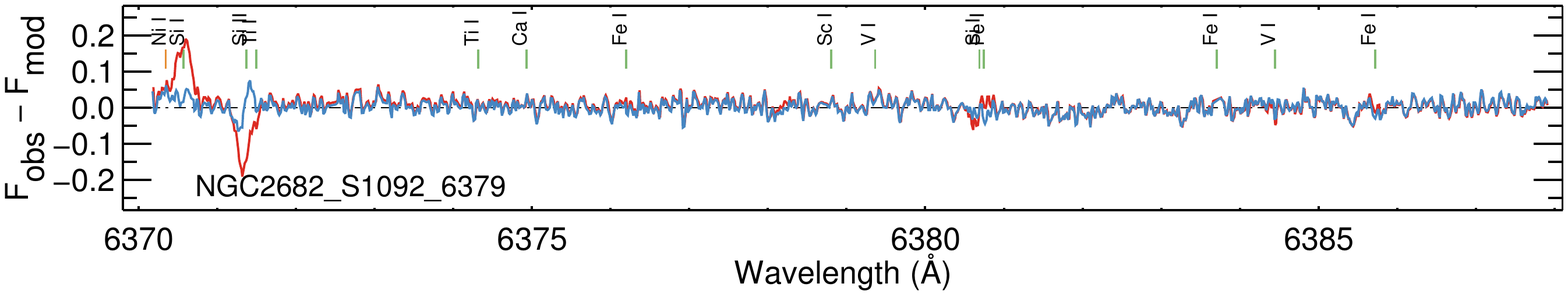}
   \includegraphics[width=0.49\hsize,  angle=180, bb=5 468 780 572, clip=]{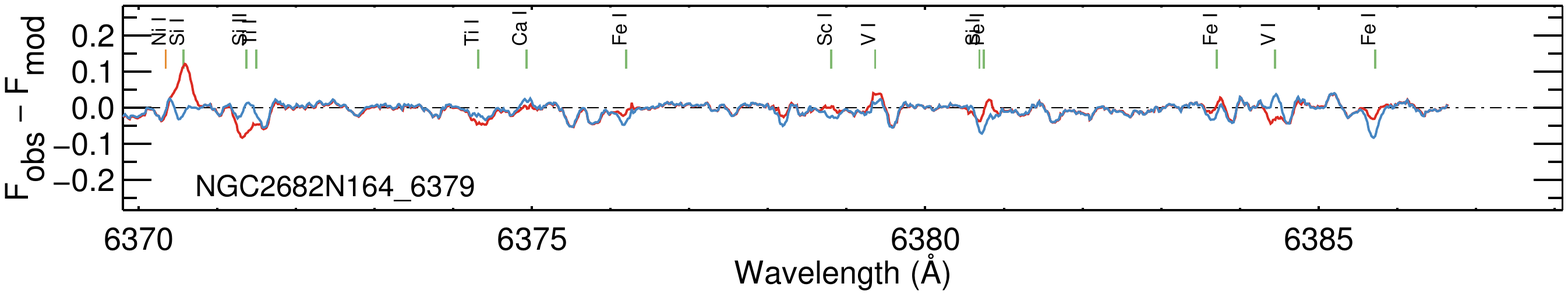}\\
   \includegraphics[width=0.49\hsize,  angle=180, bb=5 468 780 572, clip=]{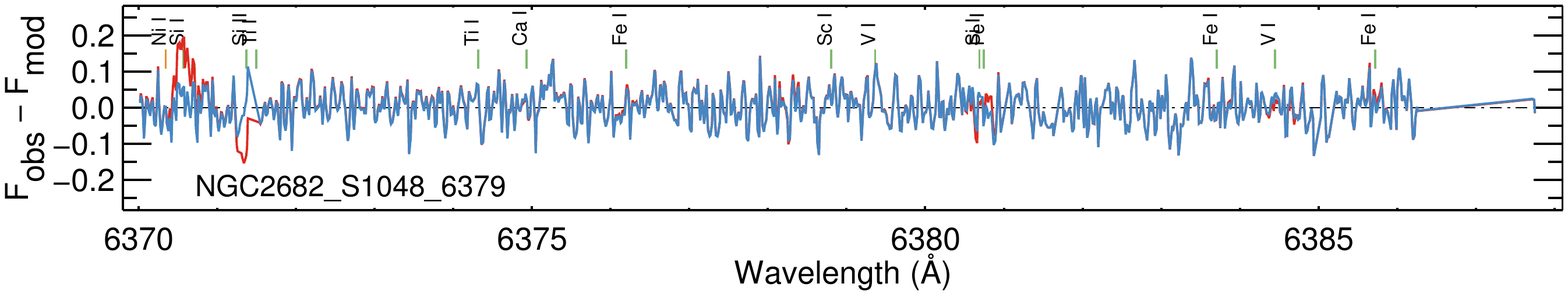}
   \includegraphics[width=0.49\hsize,  angle=180, bb=5 468 780 572, clip=]{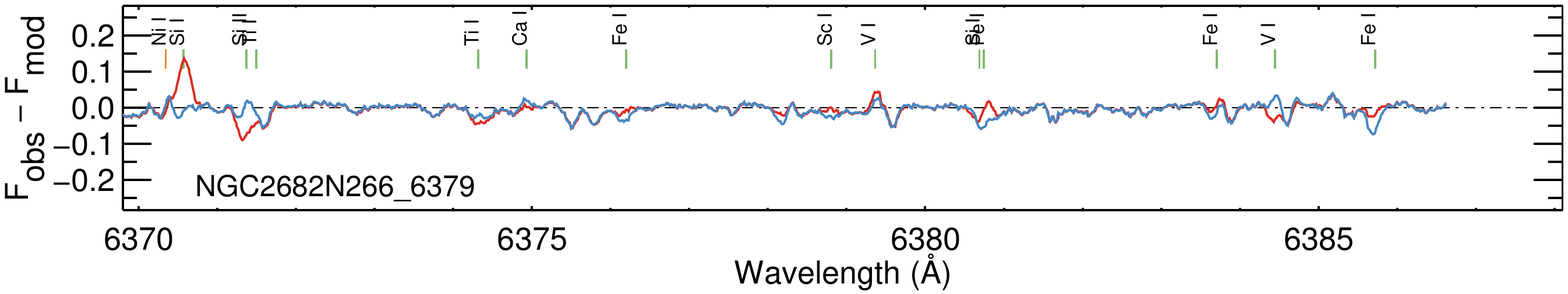}\\
   \includegraphics[width=0.49\hsize,  angle=180, bb=5 468 780 572, clip=]{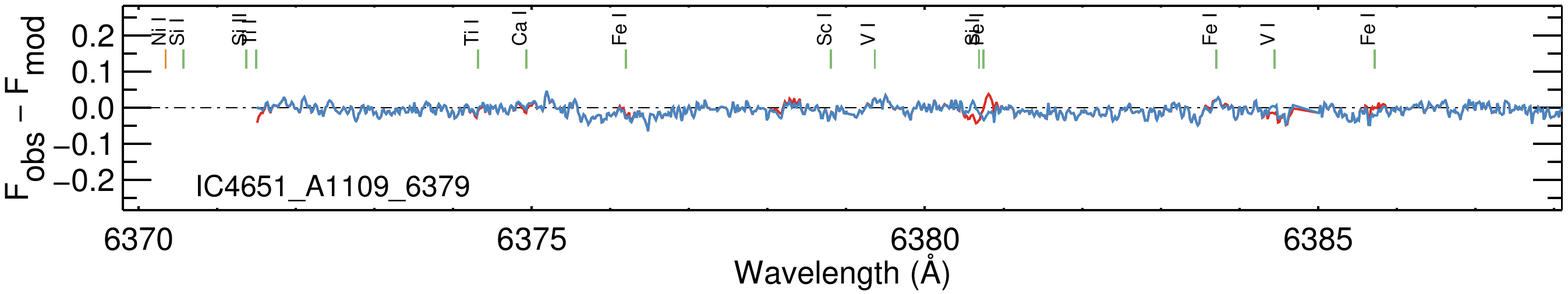}
   \includegraphics[width=0.49\hsize,  angle=180, bb=5 468 780 572, clip=]{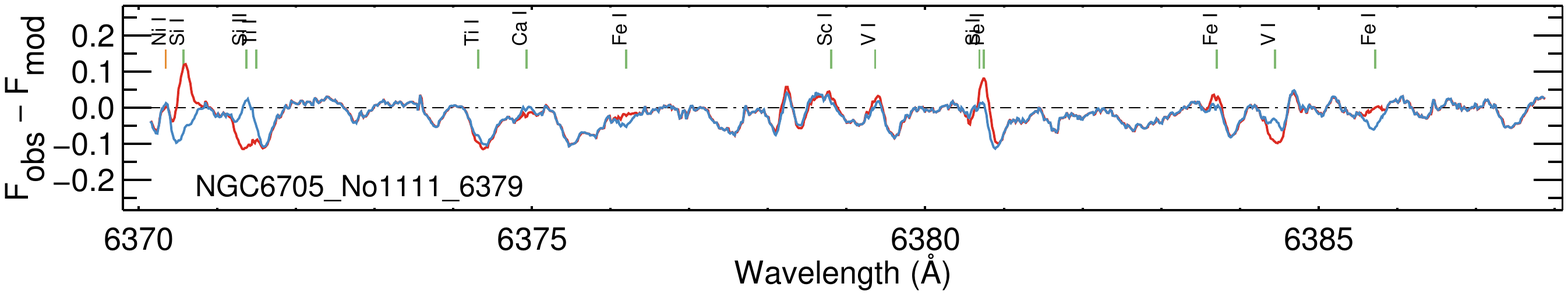}\\
   \includegraphics[width=0.49\hsize,  angle=180, bb=5 468 780 612, clip=]{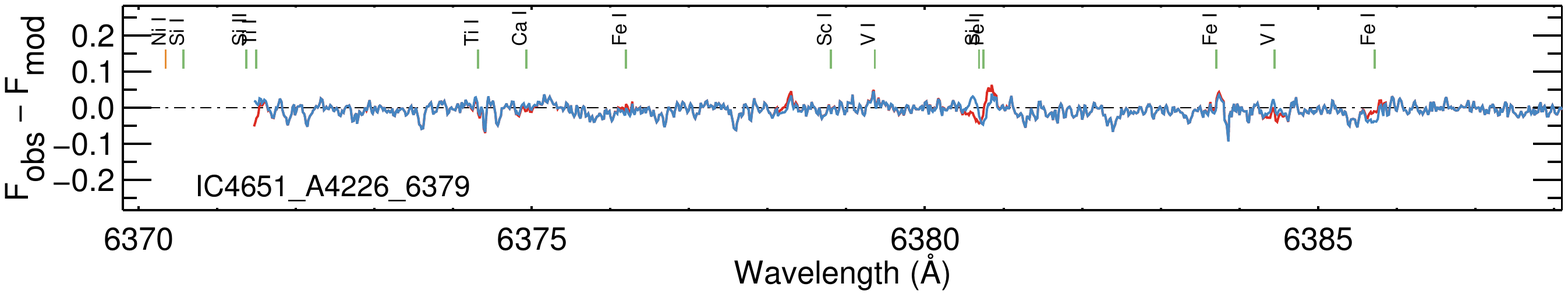}
   \includegraphics[width=0.49\hsize,  angle=180, bb=5 468 780 612, clip=]{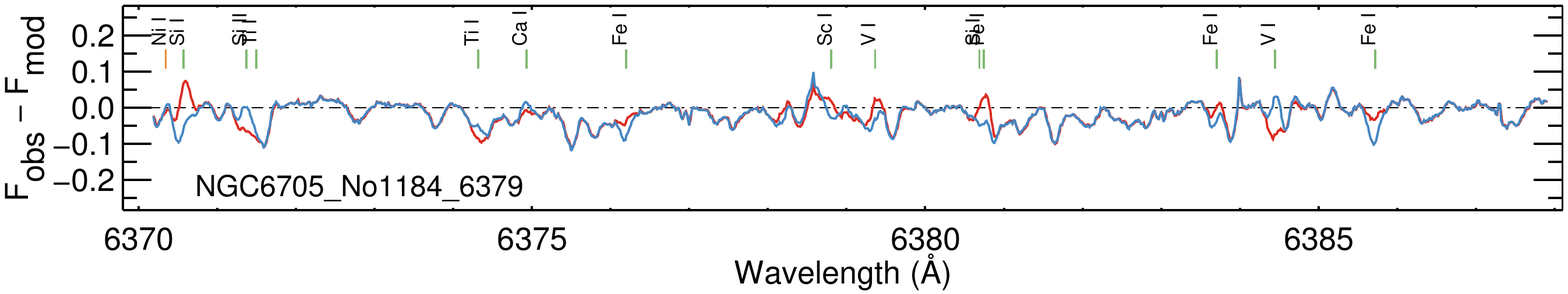}
 \caption{Same  as for Fig. \ref{figresi6196}  but  for the DIB at $\lambda$6379.3~\AA. }
\label{figresi6379}
 \end{figure*}

We tested our updated line lists using the UVES spectra presented in Sec. \ref{sectestspectra}.  A thorough description of the strategy to measure the DIB equivalent widths can be found  in \citet{Puspitarini13} and \citet{MonrealIbero15b} and will not be repeated here. In short, the spectra are modeled as the product of a series of functions that reproduce the stellar absorption features, the stellar global continuum, the interstellar absorption (i.e. the DIB), and the telluric transmission. 
DIBs were expected to be weak, specially along the sight-lines with low reddening. Thus, additional constrains were needed to make the fits converge. Specifically, we impose limits on the expected velocities for the DIBs, based on the direction of the sky and on the results for \hi\,  \citep{Brand93,Kalberla10}.
Each UVES spectrum was fitted twice. The only brick that we changed in these two fits was the function representing the stellar absorption features. In both cases we used TURBOSPECTRUM to create the synthetic stellar spectrum but using either the original VALD-3 stellar line lists or including our updates. 
The identical stellar parameters that we used to create both models are listed in Table \ref{stellarpara}. Also, we set [$\alpha$/Fe]=0.0 and we assumed a plane-parallel  geometry  for the dwarf stars, and a spherical geometry  for the giants.

The derived DIB parameters, both equivalent width and velocity, for the eight test spectra are presented in Table \ref{resultstestars}.
Reported errors for the equivalent widths take into account two components. The first one is the formal 1-sigma statistical errors associated to the fit. The second one is an estimation of the 
uncertainty associated to the reported stellar parameters ($T_{eff}$, $\log g$, and metallicity). For that, in those cases where the band could be strongly blended with a strong stellar line, the fit was done two times: without any mask, and masking this contaminating line. The difference between the equivalent widths derived from these two fits is an estimation of the uncertainty due to that, and as such, was added to the error budget.
In those cases where the fit did not converge (i.e. the estimated DIB velocity reached the imposed limits), we estimated an upper limit for a possible DIB using the standard deviation of the residuals and the typical width of the DIB.
Regarding the reported errors in the velocities, these include \emph{only} the formal 1-sigma statistical errors associated to the fit.

The most relevant result that one can extract from Table  \ref{resultstestars} is that the $\chi^2$ is most of the times smaller when using the synthetic spectra generated with the new line list, supporting the use of the corrections provided in Sec. \ref{seclinelist}.
Besides, even in the few cases where the $\chi^2$  is comparable, lines in the proximity of the DIB are better reproduced when using the improved line list and only a few lines at the edge of the spectral range under consideration are have relatively important residuals.
This is visualized in a graphical manner in Figures \ref{figresi6196} to \ref{figresi6379} that display the residuals for the fits of the eight stars when using the original (red) and updated (blue) line lists.  These last ones are $\lsim0.1$ for the dwarf star spectra. Regarding the giant stars, only for a few lines in the spectral range of the DIB at $\lambda$6196.0 they are slightly larger than 0.1. These lines are particularly susceptible to further improvements and have been marked in Table \ref{modifiedlines6186_6214} with an asterisk.
The determination of the $\log(gf)$ and the central wavelength can change depending on the stars used for calibration \citep[e.g. convection inside the star may have an effect on the  central wavelength of a given stellar line, see][]{Molaro12}. Thus, small additional adjustments in both $\log(gf)$ and the central wavelength using additional calibration stars can further improve the modeling of these lines.

Not every DIB was  detected in every spectrum. This is not surprising, given the covered range of stellar extinctions (see Tab. \ref{testspectra}). The DIB that was detected in a larger number of spectra is that at $\lambda$6283.8. The derived equivalent widths are smaller when using the up-dated line list in all cases but IC4651AM4226. This is very much in line with the result by \citet{Kohl16} for the DIB at $\lambda$5780 who, after carefully quantifying the stellar absorption in some spectra with previously reported ISM absorption, did not find hints of DIB detection and highlight the importance of adequately reproducing the stellar spectrum.

\begin{figure*}[!th] \centering
   \includegraphics[width=0.312\hsize, angle=90,  bb=0 195 270 610, clip=]{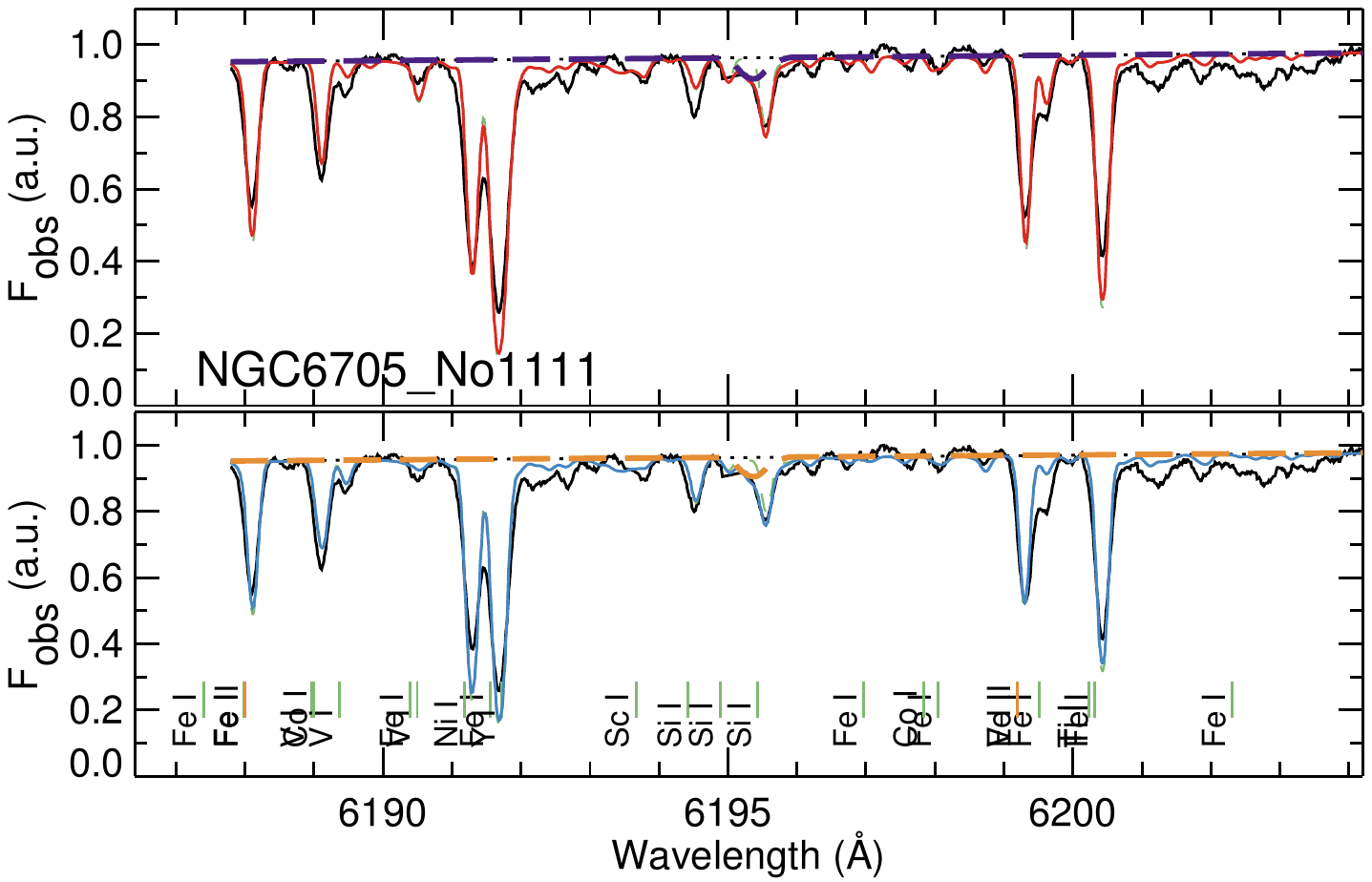}
   \includegraphics[width=0.312\hsize, angle=90,  bb=0 195 270 610, clip=]{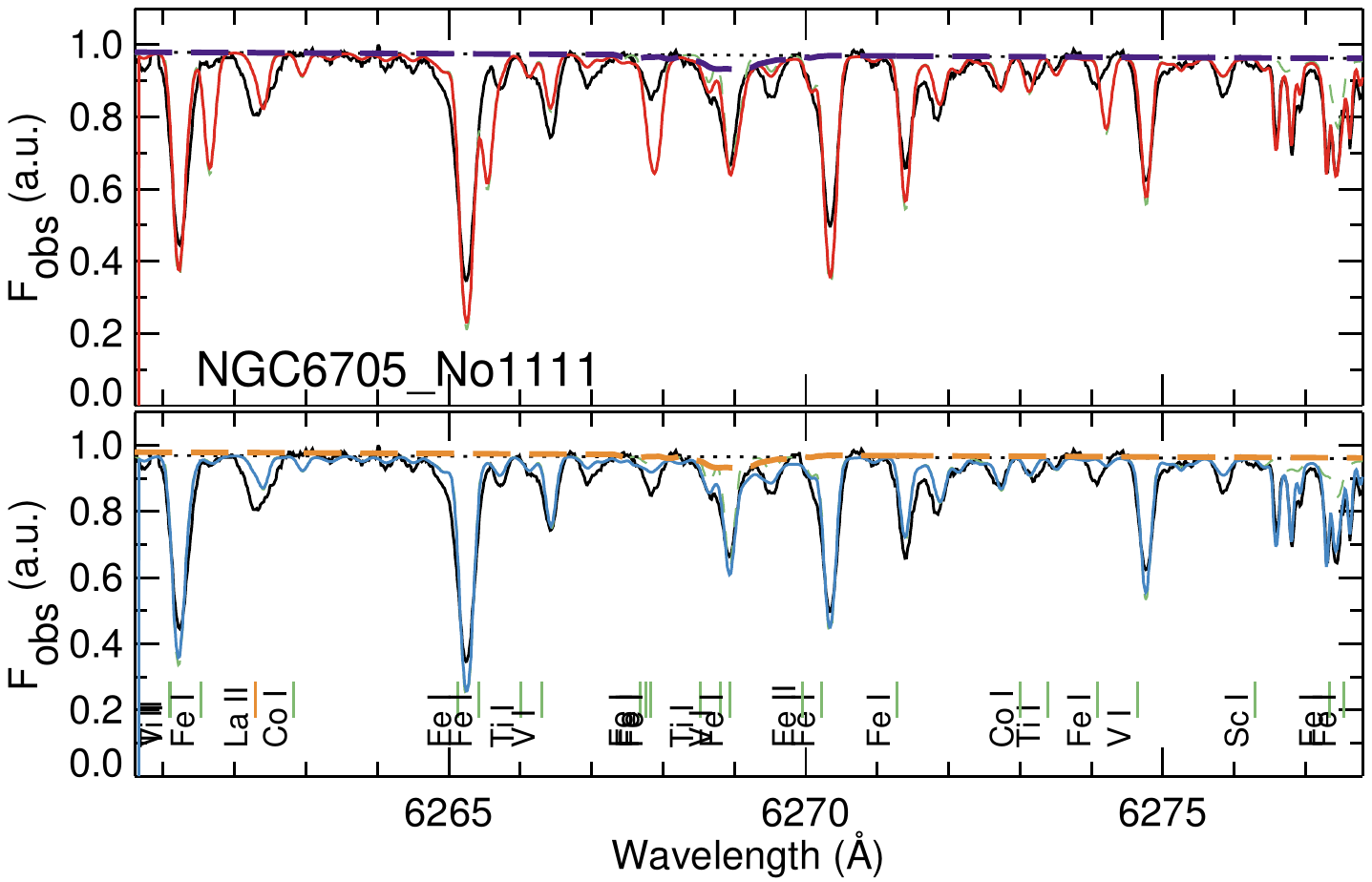}
   \includegraphics[width=0.312\hsize, angle=90,  bb=0 195 270 610, clip=]{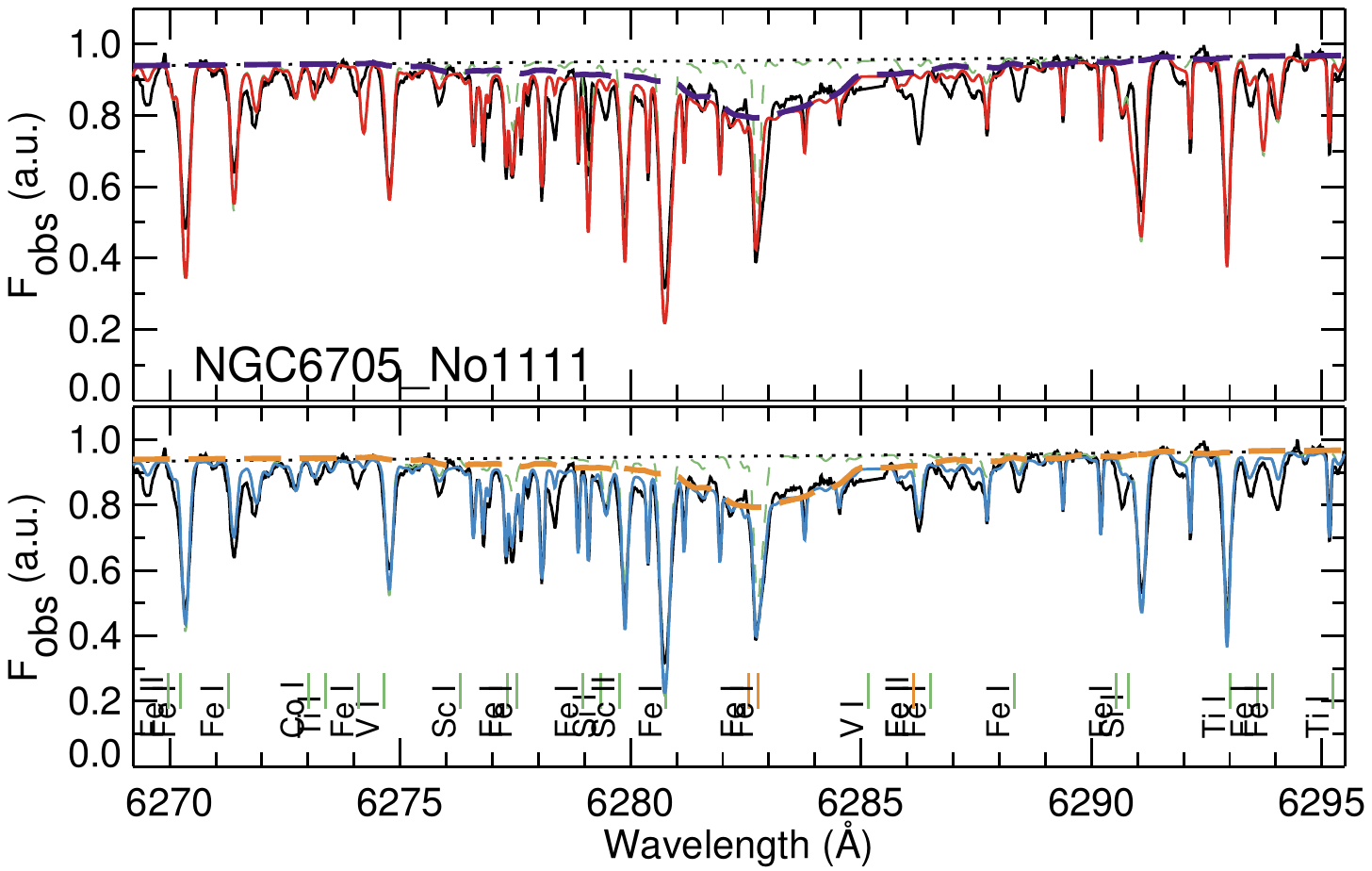}
   \includegraphics[width=0.312\hsize, angle=90,  bb=0 195 270 610, clip=]{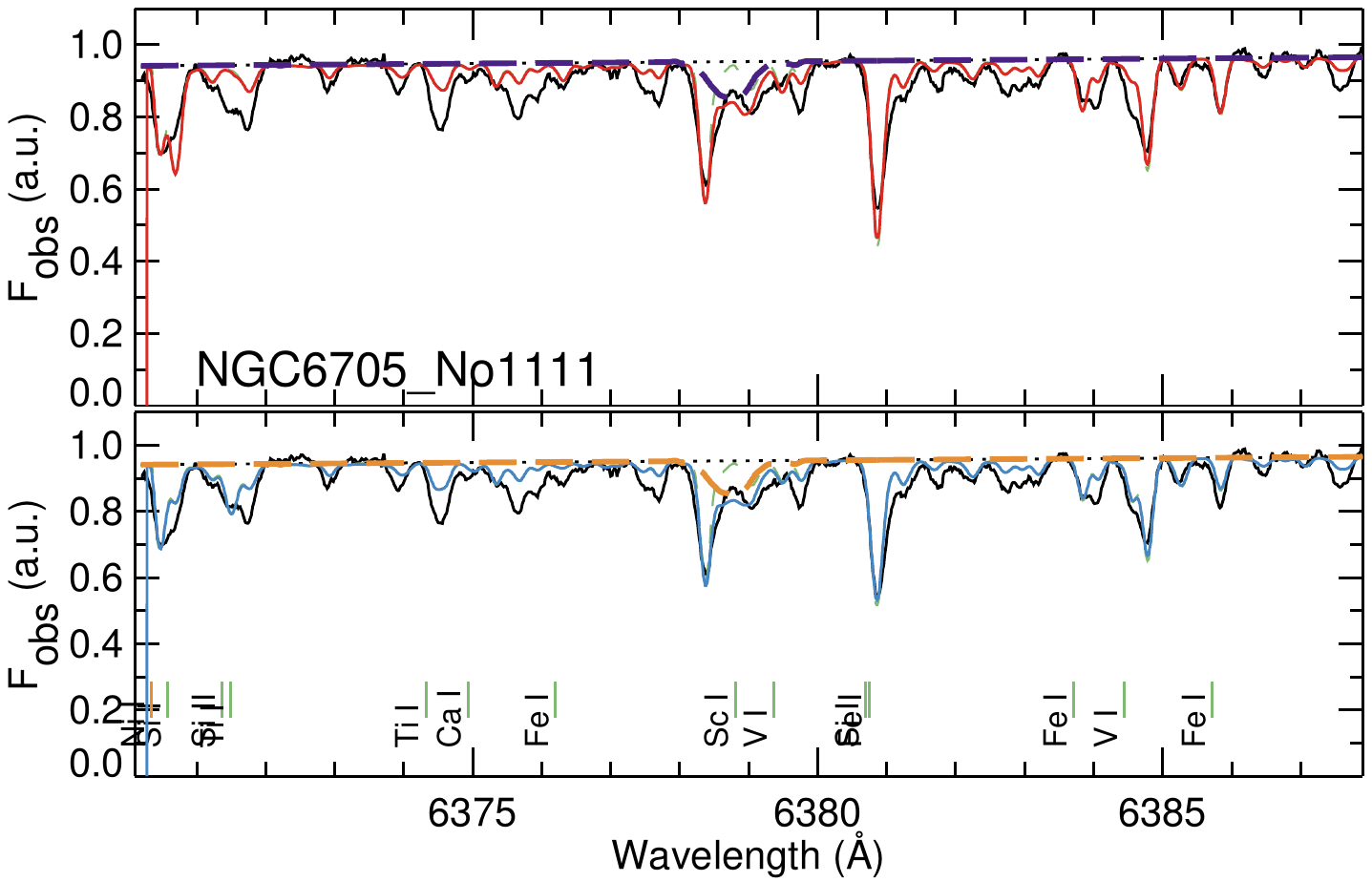}
 \caption{NGC6705\_No1111 as example of fit in the spectral region for each DIB. Total spectra are represented with solid lines as follows: black - observed spectra; red - fit with original line list; blue - fit with updated line list. Components associated to the modeled DIBs are displayed with long-dashed violet and orange  lines while the synthetic spectra as created by TURBOSPECTRUM appear as green short-dashed lines. }
\label{examplefit}
 \end{figure*}

The other DIBs were only detected in some of the stars. To have an idea of the relative importance of all the different components playing a role (different ISM absorptions, stellar spectrum, etc.) we show the fits for the four DIBs in NGC6705\_No1111  in Figure \ref{examplefit}. Note that this is both a giant and a relatively high-metallicity star. This implies that stellar absorptions are particularly strong here, and thus this is representative of one of the most difficult examples that one might find in a putative future exploitation of MOS spectra for ISM studies.
All these points are equally applicable to the other star in the cluster, NGC6705\_No1184.
In particular, the fitting algorithm was able to identify an excess of absorption at the position for the DIB at $\lambda$6269. However, this line is strongly blended with three stellar lines at $\sim\lambda6268$ (\ion{Ti}{I}, \ion{V}{I}, \ion{Fe}{i}), which might prevent an accurate determination of the DIB centroid (i.e. velocity) and explain the differences in velocity with respect to he DIB at $\lambda$6283.8.
Alternatively, small uncertainties in the automatic reduction of the archive data by the ESO pipeline (e.g. local issues with flat-fielding, fringing) could play a role in the determination of the centroid of large features, as is the case of the DIB at $\lambda$6283.8. A final source of uncertainty, regarding the velocities, is  the structure in velocity for the ISM itself along the line of sight. Since our goal here is testing the line list, not the fitting technique, we used the simple approach of reproducing the ISM absorptions by one single component (i.e. one cloud). However, this is not necessarily true. For example, in the direction of \object{NGC\,6705}, \hi\, observations show at least the existence of two clouds at very different velocities \citep{Kalberla10}. The simplicity of a fit using only one component coupled with differences in the physical conditions between these clouds might also explain the velocities differences observed between the different DIBs.

All in all, the line list corrections provided here offer the possibility of making better use of the colossal amount of cold star spectra generated in the forthcoming years  for DIB research. From the technical point of view, the most promising DIB, is the one at $\lambda$6283.8. Armed with a good fitting algorithm, an accurate synthetic stellar spectrum (as the one that can be generated using this updated line list) and appropriate tools for good telluric corrections, like TAPAS here, but also e.g. Molecfit \citep{Smette15}, researchers will be able to extract plenty of material that will help to constrain the structure of the Galactic ISM. However, given the dependence of this DIB with the environment, additional information coming from e.g. other DIBs will be desirable, and certainly feasible in the light of sights with higher extinction.

\section{Summary and conclusions  \label{secconclu}}

In this paper we improve the extraction of the parameters associated to the DIBs in the spectra of cool stars by providing 
updated line lists in three spectral ranges associated the four strong DIBs at $\lambda$6196.0,   $\lambda$6269.8, $\lambda$6283.8, and $\lambda$6379.3.
For that we tuned the $\log (gf)$ (and occasionally the central wavelength) of the stellar lines and created synthetic spectra with TURBOSPECTRUM for the Sun and Arcturus, that minimized the residuals for both stars when comparing synthetic and observed spectra.

The final stellar synthetic spectra with the improved line lists reproduce the observed spectra for these stars with greater accuracy.
The global standard deviation for the residual in a given spectral range when using TURBOSPECTRUM with the updated line lists are typically about half of those using the original lists,  and with
local residuals \lsim10\%, for both stars. 
We tested our updated line list with a set of UVES spectra for eight stars both, dwarf and giants, and suffering from different amount of extinction. The quality of the fit, as characterized by the $\chi^2$, was better when using our updated line lists, and thus we offer them to the community.

Despite the general improvement of the models, discrepancies between the radial velocities measured for the $\lambda$6283.8 DIB and for the other DIBs remain. However, based on a similar fitting method radial velocities were consistently measured in the spectra of the ESO-Gaia Survey at high optical extinctions  \citep[$A_V\sim1$~ mag, see Figs. 8, 9, 12, 14 and 16 by][]{Puspitarini15}. The examples chosen here (lower extinction towards metallic giant stars) are much more challenging and our results strongly suggest that additional work needs to be done either to improve the updated atomic data, and/or the profiles of the theoretical stellar lines to reach the high degree of precision (a few percent) that will allow to derive reliable radial velocity in these more challenging cases.

Here, we have focused on the improvement of the stellar modeling in three specific spectral ranges, corresponding to four strong DIBs. %
Inaccuracies in the stellar atmospheric modeling have already identified in other spectral ranges of interest for ISM research, as the one associated to the DIBs at $\lambda$5780 and $\lambda$5797  \citep{Kohl16}.
Given the value of the ratio between these two DIBs  as a tracer of the characteristics of the environment (i.e. radiation field) \citep{Cami97,Vos11,Cordiner13}, there is a strong motivation for a similar work to the one presented here for this spectral range. 
Likewise, the DIB at $\lambda$6614 is relatively strong and, as such, the creation of synthetic spectra in this spectral range should be contemplated, though  a preliminary exploration of this range, not discussed here, points towards a need for a more delicate approach, where for some stellar lines, additional parameters other than $\log(gf)$ should be tuned and/or effects like departures form LTE, explored.
These updated line lists together with the enormous data base that will be generated in the forthcoming years will offer an extraordinary opportunity to learn about the nature, and statistical properties of DIBs. Likewise, their suitability as probes of the properties of the ISM (e.g. distribution, excitation, etc.) will be significantly enhanced.
In particular, the assigned value for the foreground extinction affects the determination of the stellar parameters (e.g. effective temperature, metallicity).
DIBs, with their capacity to act as independent estimators of the extinction, will play a fundamental role in constraining this value.  

\begin{acknowledgements}
We thank Piercarlo Bonifacio and Elisabetta Caffau for useful advices about the focus of this work and the use of TURBOSPECTRUM.
We also thank Monique Spite for allowing us to use her piece of code that puts the Vald-3 line lists in a format adequate for TURBOSPECTRUM and with whom we shared stimulating discussions regarding the work presented here.
We also thank the referee for the useful comments that have significantly improved the first submitted version of this paper. 
%
We acknowledge support from Agence Nationale de
la Recherche through the STILISM project (ANR-12-BS05-0016-02).
%
This work has made use of the VALD database, operated at Uppsala University, the Institute of Astronomy RAS in Moscow, and the University of Vienna.
This research has made use of the NASA/IPAC Extragalactic Database (NED) which is operated by the Jet Propulsion Laboratory, California Institute of Technology, under contract with the National Aeronautics and Space Administration.

\end{acknowledgements}

\bibliography{mybib_aa}{} \bibliographystyle{./aa}

\begin{thebibliography}{79}
\expandafter\ifx\csname natexlab\endcsname\relax\def\natexlab#1{#1}\fi

\bibitem[{{Alvarez} \& {Plez}(1998)}]{Alvarez98}
{Alvarez}, R. \& {Plez}, B. 1998, \aap, 330, 1109

\bibitem[{{Bagnulo} {et~al.}(2003){Bagnulo}, {Jehin}, {Ledoux}, {Cabanac},
  {Melo}, {Gilmozzi}, \& {ESO Paranal Science Operations Team}}]{Bagnulo03}
{Bagnulo}, S., {Jehin}, E., {Ledoux}, C., {et~al.} 2003, The Messenger, 114, 10

\bibitem[{{Baron} {et~al.}(2014){Baron}, {Poznanski}, {Watson}, {Yao}, \&
  {Prochaska}}]{Baron14}
{Baron}, D., {Poznanski}, D., {Watson}, D., {Yao}, Y., \& {Prochaska}, J.~X.
  2014, ArXiv e-prints

\bibitem[{{Baron} {et~al.}(2015){Baron}, {Poznanski}, {Watson}, {Yao}, \&
  {Prochaska}}]{Baron15a}
{Baron}, D., {Poznanski}, D., {Watson}, D., {Yao}, Y., \& {Prochaska}, J.~X.
  2015, \mnras, 447, 545

\bibitem[{{Bertaux} {et~al.}(2014){Bertaux}, {Lallement}, {Ferron}, {Boonne},
  \& {Bodichon}}]{Bertaux14}
{Bertaux}, J.~L., {Lallement}, R., {Ferron}, S., {Boonne}, C., \& {Bodichon},
  R. 2014, \aap, 564, A46

\bibitem[{{Bhatt} \& {Cami}(2015)}]{Bhatt15}
{Bhatt}, N.~H. \& {Cami}, J. 2015, \apjs, 216, 22

\bibitem[{{Brand} \& {Blitz}(1993)}]{Brand93}
{Brand}, J. \& {Blitz}, L. 1993, \aap, 275, 67

\bibitem[{{Cami} {et~al.}(1997){Cami}, {Sonnentrucker}, {Ehrenfreund}, \&
  {Foing}}]{Cami97}
{Cami}, J., {Sonnentrucker}, P., {Ehrenfreund}, P., \& {Foing}, B.~H. 1997,
  \aap, 326, 822

\bibitem[{{Campbell} {et~al.}(2015){Campbell}, {Holz}, {Gerlich}, \&
  {Maier}}]{Campbell15}
{Campbell}, E.~K., {Holz}, M., {Gerlich}, D., \& {Maier}, J.~P. 2015, \nat,
  523, 322

\bibitem[{{Chen} {et~al.}(2013){Chen}, {Lallement}, {Babusiaux}, {Puspitarini},
  {Bonifacio}, \& {Hill}}]{Chen13}
{Chen}, H.-C., {Lallement}, R., {Babusiaux}, C., {et~al.} 2013, \aap, 550, A62

\bibitem[{{Cirasuolo} {et~al.}(2014){Cirasuolo}, {Afonso}, {Carollo}, {Flores},
  {Maiolino}, {Oliva}, {Paltani}, {Vanzi}, {Evans}, {Abreu}, {Atkinson},
  {Babusiaux}, {Beard}, {Bauer}, {Bellazzini}, {Bender}, {Best}, {Bezawada},
  {Bonifacio}, {Bragaglia}, {Bryson}, {Busher}, {Cabral}, {Caputi}, {Centrone},
  {Chemla}, {Cimatti}, {Cioni}, {Clementini}, {Coelho}, {Crnojevic}, {Daddi},
  {Dunlop}, {Eales}, {Feltzing}, {Ferguson}, {Fisher}, {Fontana}, {Fynbo},
  {Garilli}, {Gilmore}, {Glauser}, {Guinouard}, {Hammer}, {Hastings}, {Hess},
  {Ivison}, {Jagourel}, {Jarvis}, {Kaper}, {Kauffman}, {Kitching}, {Lawrence},
  {Lee}, {Lemasle}, {Licausi}, {Lilly}, {Lorenzetti}, {Lunney}, {Maiolino},
  {Mannucci}, {McLure}, {Minniti}, {Montgomery}, {Muschielok}, {Nandra},
  {Navarro}, {Norberg}, {Oliver}, {Origlia}, {Padilla}, {Peacock}, {Pedichini},
  {Peng}, {Pentericci}, {Pragt}, {Puech}, {Randich}, {Rees}, {Renzini}, {Ryde},
  {Rodrigues}, {Roseboom}, {Royer}, {Saglia}, {Sanchez}, {Schiavon},
  {Schnetler}, {Sobral}, {Speziali}, {Sun}, {Stuik}, {Taylor}, {Taylor},
  {Todd}, {Tolstoy}, {Torres}, {Tosi}, {Vanzella}, {Venema}, {Vitali},
  {Wegner}, {Wells}, {Wild}, {Wright}, {Zamorani}, \& {Zoccali}}]{Cirasuolo14}
{Cirasuolo}, M., {Afonso}, J., {Carollo}, M., {et~al.} 2014, in \procspie, Vol.
  9147, Ground-based and Airborne Instrumentation for Astronomy V, 91470N

\bibitem[{{Cordiner} {et~al.}(2011){Cordiner}, {Cox}, {Evans}, {Trundle},
  {Smith}, {Sarre}, \& {Gordon}}]{Cordiner11}
{Cordiner}, M.~A., {Cox}, N.~L.~J., {Evans}, C.~J., {et~al.} 2011, \apj, 726,
  39

\bibitem[{{Cordiner} {et~al.}(2013){Cordiner}, {Fossey}, {Smith}, \&
  {Sarre}}]{Cordiner13}
{Cordiner}, M.~A., {Fossey}, S.~J., {Smith}, A.~M., \& {Sarre}, P.~J. 2013,
  \apjl, 764, L10

\bibitem[{{Cordiner} {et~al.}(2008){Cordiner}, {Smith}, {Cox}, {Evans},
  {Hunter}, {Przybilla}, {Bresolin}, \& {Sarre}}]{Cordiner08}
{Cordiner}, M.~A., {Smith}, K.~T., {Cox}, N.~L.~J., {et~al.} 2008, \aap, 492,
  L5

\bibitem[{{Cox} {et~al.}(2014){Cox}, {Cami}, {Kaper}, {Ehrenfreund}, {Foing},
  {Ochsendorf}, {van Hooff}, \& {Salama}}]{Cox14}
{Cox}, N.~L.~J., {Cami}, J., {Kaper}, L., {et~al.} 2014, \aap, 569, A117

\bibitem[{{Cox} {et~al.}(2007){Cox}, {Cordiner}, {Ehrenfreund}, {Kaper},
  {Sarre}, {Foing}, {Spaans}, {Cami}, {Sofia}, {Clayton}, {Gordon}, \&
  {Salama}}]{Cox07}
{Cox}, N.~L.~J., {Cordiner}, M.~A., {Ehrenfreund}, P., {et~al.} 2007, \aap,
  470, 941

\bibitem[{{Dahlstrom} {et~al.}(2013){Dahlstrom}, {York}, {Welty}, {Oka},
  {Hobbs}, {Johnson}, {Friedman}, {Jiang}, {Rachford}, {Sherman}, {Snow}, \&
  {Sonnentrucker}}]{Dahlstrom13}
{Dahlstrom}, J., {York}, D.~G., {Welty}, D.~E., {et~al.} 2013, \apj, 773, 41

\bibitem[{{Dalton} {et~al.}(2014){Dalton}, {Trager}, {Abrams}, {Bonifacio},
  {L{\'o}pez Aguerri}, {Middleton}, {Benn}, {Dee}, {Say{\`e}de}, {Lewis},
  {Pragt}, {Pico}, {Walton}, {Rey}, {Allende Prieto}, {Pe{\~n}ate}, {Lhome},
  {Ag{\'o}cs}, {Alonso}, {Terrett}, {Brock}, {Gilbert}, {Ridings}, {Guinouard},
  {Verheijen}, {Tosh}, {Rogers}, {Steele}, {Stuik}, {Tromp}, {Jasko}, {Kragt},
  {Lesman}, {Mottram}, {Bates}, {Gribbin}, {Rodriguez}, {Delgado}, {Martin},
  {Cano}, {Navarro}, {Irwin}, {Lewis}, {Gonzalez Solares}, {O'Mahony},
  {Bianco}, {Zurita}, {ter Horst}, {Molinari}, {Lodi}, {Guerra}, {Vallenari},
  \& {Baruffolo}}]{Dalton14}
{Dalton}, G., {Trager}, S., {Abrams}, D.~C., {et~al.} 2014, in \procspie, Vol.
  9147, Ground-based and Airborne Instrumentation for Astronomy V, 91470L

\bibitem[{{de Jong} {et~al.}(2014){de Jong}, {Barden}, {Bellido-Tirado},
  {Brynnel}, {Chiappini}, {Depagne}, {Haynes}, {Johl}, {Phillips}, {Schnurr},
  {Schwope}, {Walcher}, {Bauer}, {Cescutti}, {Cioni}, {Dionies}, {Enke},
  {Haynes}, {Kelz}, {Kitaura}, {Lamer}, {Minchev}, {M{\"u}ller}, {Nuza},
  {Olaya}, {Piffl}, {Popow}, {Saviauk}, {Steinmetz}, {Ural}, {Valentini},
  {Winkler}, {Wisotzki}, {Ansorge}, {Banerji}, {Gonzalez Solares}, {Irwin},
  {Kennicutt}, {King}, {McMahon}, {Koposov}, {Parry}, {Sun}, {Walton},
  {Finger}, {Iwert}, {Krumpe}, {Lizon}, {Mainieri}, {Amans}, {Bonifacio},
  {Cohen}, {Fran{\c c}ois}, {Jagourel}, {Mignot}, {Royer}, {Sartoretti},
  {Bender}, {Hess}, {Lang-Bardl}, {Muschielok}, {Schlichter}, {B{\"o}hringer},
  {Boller}, {Bongiorno}, {Brusa}, {Dwelly}, {Merloni}, {Nandra}, {Salvato},
  {Pragt}, {Navarro}, {Gerlofsma}, {Roelfsema}, {Dalton}, {Middleton}, {Tosh},
  {Boeche}, {Caffau}, {Christlieb}, {Grebel}, {Hansen}, {Koch}, {Ludwig},
  {Mandel}, {Quirrenbach}, {Sbordone}, {Seifert}, {Thimm}, {Helmi}, {trager},
  {Bensby}, {Feltzing}, {Ruchti}, {Edvardsson}, {Korn}, {Lind}, {Boland},
  {Colless}, {Frost}, {Gilbert}, {Gillingham}, {Lawrence}, {Legg}, {Saunders},
  {Sheinis}, {Driver}, {Robotham}, {Bacon}, {Caillier}, {Kosmalski}, {Laurent},
  \& {Richard}}]{deJong14}
{de Jong}, R.~S., {Barden}, S., {Bellido-Tirado}, O., {et~al.} 2014, in
  \procspie, Vol. 9147, Ground-based and Airborne Instrumentation for Astronomy
  V, 91470M

\bibitem[{{Dekker} {et~al.}(2000){Dekker}, {D'Odorico}, {Kaufer}, {Delabre}, \&
  {Kotzlowski}}]{Dekker00}
{Dekker}, H., {D'Odorico}, S., {Kaufer}, A., {Delabre}, B., \& {Kotzlowski}, H.
  2000, in Society of Photo-Optical Instrumentation Engineers (SPIE) Conference
  Series, Vol. 4008, Optical and IR Telescope Instrumentation and Detectors,
  ed. M.~{Iye} \& A.~F. {Moorwood}, 534--545

\bibitem[{{Ehrenfreund} {et~al.}(2002){Ehrenfreund}, {Cami},
  {Jim{\'e}nez-Vicente}, {Foing}, {Kaper}, {van der Meer}, {Cox},
  {D'Hendecourt}, {Maier}, {Salama}, {Sarre}, {Snow}, \&
  {Sonnentrucker}}]{Ehrenfreund02}
{Ehrenfreund}, P., {Cami}, J., {Jim{\'e}nez-Vicente}, J., {et~al.} 2002, \apjl,
  576, L117

\bibitem[{{Elyajouri} {et~al.}(2016){Elyajouri}, {Monreal-Ibero}, {Remy}, \&
  {Lallement}}]{Elyajouri16}
{Elyajouri}, M., {Monreal-Ibero}, A., {Remy}, Q., \& {Lallement}, R. 2016,
  \apjs, 225, 19

\bibitem[{{Foing} \& {Ehrenfreund}(1994)}]{Foing94}
{Foing}, B.~H. \& {Ehrenfreund}, P. 1994, \nat, 369, 296

\bibitem[{{Friedman} {et~al.}(2011){Friedman}, {York}, {McCall}, {Dahlstrom},
  {Sonnentrucker}, {Welty}, {Drosback}, {Hobbs}, {Rachford}, \&
  {Snow}}]{Friedman11}
{Friedman}, S.~D., {York}, D.~G., {McCall}, B.~J., {et~al.} 2011, \apj, 727, 33

\bibitem[{{Fulara} \& {Kre{\l}owski}(2000)}]{Fulara00}
{Fulara}, J. \& {Kre{\l}owski}, J. 2000, \nar, 44, 581

\bibitem[{{Galazutdinov} {et~al.}(2000){Galazutdinov}, {Musaev},
  {Kre{\l}owski}, \& {Walker}}]{Galazutdinov00}
{Galazutdinov}, G.~A., {Musaev}, F.~A., {Kre{\l}owski}, J., \& {Walker},
  G.~A.~H. 2000, \pasp, 112, 648

\bibitem[{{Gustafsson} {et~al.}(2008){Gustafsson}, {Edvardsson}, {Eriksson},
  {J{\o}rgensen}, {Nordlund}, \& {Plez}}]{Gustafsson08}
{Gustafsson}, B., {Edvardsson}, B., {Eriksson}, K., {et~al.} 2008, \aap, 486,
  951

\bibitem[{{Hamano} {et~al.}(2016){Hamano}, {Kobayashi}, {Kondo}, {Sameshima},
  {Nakanishi}, {Ikeda}, {Yasui}, {Mizumoto}, {Matsunaga}, {Fukue}, {Yamamoto},
  {Izumi}, {Mito}, {Nakaoka}, {Kawanishi}, {Kitano}, {Otsubo}, {Kinoshita}, \&
  {Kawakita}}]{Hamano16}
{Hamano}, S., {Kobayashi}, N., {Kondo}, S., {et~al.} 2016, \apj, 821, 42

\bibitem[{{Heckman} \& {Lehnert}(2000)}]{Heckman00}
{Heckman}, T.~M. \& {Lehnert}, M.~D. 2000, \apj, 537, 690

\bibitem[{{Heger}(1922)}]{heg22}
{Heger}, M.~L. 1922, Lick Observatory Bulletin, 10, 141

\bibitem[{{Herbig}(1993)}]{Herbig93}
{Herbig}, G.~H. 1993, \apj, 407, 142

\bibitem[{{Herbig}(1995)}]{her95}
{Herbig}, G.~H. 1995, \araa, 33, 19

\bibitem[{{Hobbs} {et~al.}(2008){Hobbs}, {York}, {Snow}, {Oka}, {Thorburn},
  {Bishof}, {Friedman}, {McCall}, {Rachford}, {Sonnentrucker}, \&
  {Welty}}]{Hobbs08}
{Hobbs}, L.~M., {York}, D.~G., {Snow}, T.~P., {et~al.} 2008, \apj, 680, 1256

\bibitem[{{Hobbs} {et~al.}(2009){Hobbs}, {York}, {Thorburn}, {Snow}, {Bishof},
  {Friedman}, {McCall}, {Oka}, {Rachford}, {Sonnentrucker}, \& {Welty}}]{hob09}
{Hobbs}, L.~M., {York}, D.~G., {Thorburn}, J.~A., {et~al.} 2009, \apj, 705, 32

\bibitem[{{Iglesias-Groth}(2007)}]{igl07}
{Iglesias-Groth}, S. 2007, \apjl, 661, L167

\bibitem[{{Joblin} {et~al.}(1990){Joblin}, {D'Hendecourt}, {Leger}, \&
  {Maillard}}]{Joblin90}
{Joblin}, C., {D'Hendecourt}, L., {Leger}, A., \& {Maillard}, J.~P. 1990, \nat,
  346, 729

\bibitem[{{Jofr{\'e}} {et~al.}(2014){Jofr{\'e}}, {Heiter}, {Soubiran},
  {Blanco-Cuaresma}, {Worley}, {Pancino}, {Cantat-Gaudin}, {Magrini},
  {Bergemann}, {Gonz{\'a}lez Hern{\'a}ndez}, {Hill}, {Lardo}, {de Laverny},
  {Lind}, {Masseron}, {Montes}, {Mucciarelli}, {Nordlander}, {Recio Blanco},
  {Sobeck}, {Sordo}, {Sousa}, {Tabernero}, {Vallenari}, \& {Van Eck}}]{Jofre14}
{Jofr{\'e}}, P., {Heiter}, U., {Soubiran}, C., {et~al.} 2014, \aap, 564, A133

\bibitem[{{Kalberla} {et~al.}(2010){Kalberla}, {McClure-Griffiths}, {Pisano},
  {Calabretta}, {Ford}, {Lockman}, {Staveley-Smith}, {Kerp}, {Winkel},
  {Murphy}, \& {Newton-McGee}}]{Kalberla10}
{Kalberla}, P.~M.~W., {McClure-Griffiths}, N.~M., {Pisano}, D.~J., {et~al.}
  2010, \aap, 521, A17

\bibitem[{{Katz} {et~al.}(2004){Katz}, {Munari}, {Cropper}, {Zwitter},
  {Th{\'e}venin}, {David}, {Viala}, {Crifo}, {Gomboc}, {Royer}, {Arenou},
  {Marrese}, {Sordo}, {Wilkinson}, {Vallenari}, {Turon}, {Helmi}, {Bono},
  {Perryman}, {G{\'o}mez}, {Tomasella}, {Boschi}, {Morin}, {Haywood},
  {Soubiran}, {Castelli}, {Bijaoui}, {Bertelli}, {Prsa}, {Mignot}, {Sellier},
  {Baylac}, {Lebreton}, {Jauregi}, {Siviero}, {Bingham}, {Chemla}, {Coker},
  {Dibbens}, {Hancock}, {Holland}, {Horville}, {Huet}, {Laporte}, {Melse},
  {Say{\`e}de}, {Stevenson}, {Vola}, {Walton}, \& {Winter}}]{Katz04}
{Katz}, D., {Munari}, U., {Cropper}, M., {et~al.} 2004, \mnras, 354, 1223

\bibitem[{{Kharchenko} {et~al.}(2005){Kharchenko}, {Piskunov}, {R{\"o}ser},
  {Schilbach}, \& {Scholz}}]{Kharchenko05}
{Kharchenko}, N.~V., {Piskunov}, A.~E., {R{\"o}ser}, S., {Schilbach}, E., \&
  {Scholz}, R.-D. 2005, \aap, 438, 1163

\bibitem[{{Kohl} {et~al.}(2016){Kohl}, {Czesla}, \& {Schmitt}}]{Kohl16}
{Kohl}, S., {Czesla}, S., \& {Schmitt}, J.~H.~M.~M. 2016, \aap, 591, A20

\bibitem[{{Kokkin} {et~al.}(2008){Kokkin}, {Troy}, {Nakajima}, {Nauta},
  {Varberg}, {Metha}, {Lucas}, \& {Schmidt}}]{Kokkin08}
{Kokkin}, D.~L., {Troy}, T.~P., {Nakajima}, M., {et~al.} 2008, \apjl, 681, L49

\bibitem[{{Kos} {et~al.}(2014){Kos}, {Zwitter}, {Wyse}, {Bienaym{\'e}},
  {Binney}, {Bland-Hawthorn}, {Freeman}, {Gibson}, {Gilmore}, {Grebel},
  {Helmi}, {Kordopatis}, {Munari}, {Navarro}, {Parker}, {Reid}, {Seabroke},
  {Sharma}, {Siebert}, {Siviero}, {Steinmetz}, {Watson}, \& {Williams}}]{Kos14}
{Kos}, J., {Zwitter}, T., {Wyse}, R., {et~al.} 2014, Science, 345, 791

\bibitem[{{Kurucz}(2005)}]{Kurucz05}
{Kurucz}, R.~L. 2005, Memorie della Societa Astronomica Italiana Supplementi,
  8, 189

\bibitem[{{Lan} {et~al.}(2015){Lan}, {M{\'e}nard}, \& {Zhu}}]{Lan15}
{Lan}, T.-W., {M{\'e}nard}, B., \& {Zhu}, G. 2015, \mnras, 452, 3629

\bibitem[{{Lawton} {et~al.}(2008){Lawton}, {Churchill}, {York}, {Ellison},
  {Snow}, {Johnson}, {Ryan}, \& {Benn}}]{Lawton08}
{Lawton}, B., {Churchill}, C.~W., {York}, B.~A., {et~al.} 2008, \aj, 136, 994

\bibitem[{{Maier} {et~al.}(2004){Maier}, {Walker}, \& {Bohlender}}]{mai04}
{Maier}, J.~P., {Walker}, G.~A.~H., \& {Bohlender}, D.~A. 2004, \apj, 602, 286

\bibitem[{{McCall} \& {Griffin}(2013)}]{mcc13}
{McCall}, B.~J. \& {Griffin}, R.~E. 2013, in Proceedings of the royal society
  A, Vol. 469, Proceedings of the royal society A, ed. {M.~Berry}, 20120604

\bibitem[{{Merrill}(1934)}]{mer34}
{Merrill}, P.~W. 1934, \pasp, 46, 206

\bibitem[{{Merrill}(1936)}]{mer36}
{Merrill}, P.~W. 1936, \apj, 83, 126

\bibitem[{{Molaro} \& {Monai}(2012)}]{Molaro12}
{Molaro}, P. \& {Monai}, S. 2012, \aap, 544, A125

\bibitem[{{Monreal-Ibero} {et~al.}(2015{\natexlab{a}}){Monreal-Ibero},
  {Lallement}, {Puspitarini}, {Bonifacio}, \& {Monaco}}]{MonrealIbero15b}
{Monreal-Ibero}, A., {Lallement}, R., {Puspitarini}, L., {Bonifacio}, P., \&
  {Monaco}, L. 2015{\natexlab{a}}, \memsai, 86, 527

\bibitem[{{Monreal-Ibero} {et~al.}(2015{\natexlab{b}}){Monreal-Ibero},
  {Weilbacher}, {Wendt}, {Selman}, {Lallement}, {Brinchmann}, {Kamann}, \&
  {Sandin}}]{MonrealIbero15}
{Monreal-Ibero}, A., {Weilbacher}, P.~M., {Wendt}, M., {et~al.}
  2015{\natexlab{b}}, \aap, 576, L3

\bibitem[{{Munari} {et~al.}(2008){Munari}, {Tomasella}, {Fiorucci},
  {Bienaym{\'e}}, {Binney}, {Bland-Hawthorn}, {Boeche}, {Campbell}, {Freeman},
  {Gibson}, {Gilmore}, {Grebel}, {Helmi}, {Navarro}, {Parker}, {Seabroke},
  {Siebert}, {Siviero}, {Steinmetz}, {Watson}, {Williams}, {Wyse}, \&
  {Zwitter}}]{Munari08}
{Munari}, U., {Tomasella}, L., {Fiorucci}, M., {et~al.} 2008, \aap, 488, 969

\bibitem[{{Pandey} {et~al.}(2010){Pandey}, {Sandhu}, {Sagar}, \&
  {Battinelli}}]{Pandey10}
{Pandey}, A.~K., {Sandhu}, T.~S., {Sagar}, R., \& {Battinelli}, P. 2010,
  \mnras, 403, 1491

\bibitem[{{Phillips} {et~al.}(2013){Phillips}, {Simon}, {Morrell}, {Burns},
  {Cox}, {Foley}, {Karakas}, {Patat}, {Sternberg}, {Williams}, {Gal-Yam},
  {Hsiao}, {Leonard}, {Persson}, {Stritzinger}, {Thompson}, {Campillay},
  {Contreras}, {Folatelli}, {Freedman}, {Hamuy}, {Roth}, {Shields}, {Suntzeff},
  {Chomiuk}, {Ivans}, {Madore}, {Penprase}, {Perley}, {Pignata}, {Preston}, \&
  {Soderberg}}]{Phillips13}
{Phillips}, M.~M., {Simon}, J.~D., {Morrell}, N., {et~al.} 2013, \apj, 779, 38

\bibitem[{{Plez}(2012)}]{Plez12}
{Plez}, B. 2012, {Turbospectrum: Code for spectral synthesis}, astrophysics
  Source Code Library

\bibitem[{{Porceddu} {et~al.}(1991){Porceddu}, {Benvenuti}, \&
  {Krelowski}}]{Porceddu91}
{Porceddu}, I., {Benvenuti}, P., \& {Krelowski}, J. 1991, \aap, 248, 188

\bibitem[{{Puspitarini} {et~al.}(2015){Puspitarini}, {Lallement}, {Babusiaux},
  {Chen}, {Bonifacio}, {Sbordone}, {Caffau}, {Duffau}, {Hill}, {Monreal-Ibero},
  {Royer}, {Arenou}, {Peralta}, {Drew}, {Bonito}, {Lopez-Santiago}, {Alfaro},
  {Bensby}, {Bragaglia}, {Flaccomio}, {Lanzafame}, {Pancino}, {Recio-Blanco},
  {Smiljanic}, {Costado}, {Lardo}, {de Laverny}, \& {Zwitter}}]{Puspitarini15}
{Puspitarini}, L., {Lallement}, R., {Babusiaux}, C., {et~al.} 2015, \aap, 573,
  A35

\bibitem[{{Puspitarini} {et~al.}(2013){Puspitarini}, {Lallement}, \&
  {Chen}}]{Puspitarini13}
{Puspitarini}, L., {Lallement}, R., \& {Chen}, H.-C. 2013, \aap, 555, A25

\bibitem[{{Raimond} {et~al.}(2012){Raimond}, {Lallement}, {Vergely},
  {Babusiaux}, \& {Eyer}}]{Raimond12}
{Raimond}, S., {Lallement}, R., {Vergely}, J.~L., {Babusiaux}, C., \& {Eyer},
  L. 2012, \aap, 544, A136

\bibitem[{{Ritchey} \& {Wallerstein}(2015)}]{Ritchey15}
{Ritchey}, A.~M. \& {Wallerstein}, G. 2015, \pasp, 127, 223

\bibitem[{{Ryabchikova} {et~al.}(2015){Ryabchikova}, {Piskunov}, {Kurucz},
  {Stempels}, {Heiter}, {Pakhomov}, \& {Barklem}}]{Ryabchikova15}
{Ryabchikova}, T., {Piskunov}, N., {Kurucz}, R.~L., {et~al.} 2015, \physscr,
  90, 054005

\bibitem[{{Salama} {et~al.}(1996){Salama}, {Bakes}, {Allamandola}, \&
  {Tielens}}]{Salama96}
{Salama}, F., {Bakes}, E.~L.~O., {Allamandola}, L.~J., \& {Tielens},
  A.~G.~G.~M. 1996, \apj, 458, 621

\bibitem[{{Santos} {et~al.}(2012){Santos}, {Lovis}, {Melendez}, {Montalto},
  {Naef}, \& {Pace}}]{Santos12}
{Santos}, N.~C., {Lovis}, C., {Melendez}, J., {et~al.} 2012, \aap, 538, A151

\bibitem[{{Santos} {et~al.}(2009){Santos}, {Lovis}, {Pace}, {Melendez}, \&
  {Naef}}]{Santos09}
{Santos}, N.~C., {Lovis}, C., {Pace}, G., {Melendez}, J., \& {Naef}, D. 2009,
  \aap, 493, 309

\bibitem[{{Sarre}(2006)}]{sar06}
{Sarre}, P.~J. 2006, Journal of Molecular Spectroscopy, 238, 1

\bibitem[{{Sassara} {et~al.}(2001){Sassara}, {Zerza}, {Chergui}, \&
  {Leach}}]{sas01}
{Sassara}, A., {Zerza}, G., {Chergui}, M., \& {Leach}, S. 2001, \apjs, 135, 263

\bibitem[{{Schlafly} \& {Finkbeiner}(2011)}]{Schlafly11}
{Schlafly}, E.~F. \& {Finkbeiner}, D.~P. 2011, \apj, 737, 103

\bibitem[{{Schlegel} {et~al.}(1998){Schlegel}, {Finkbeiner}, \&
  {Davis}}]{Schlegel98}
{Schlegel}, D.~J., {Finkbeiner}, D.~P., \& {Davis}, M. 1998, \apj, 500, 525

\bibitem[{{Smette} {et~al.}(2015){Smette}, {Sana}, {Noll}, {Horst}, {Kausch},
  {Kimeswenger}, {Barden}, {Szyszka}, {Jones}, {Gallenne}, {Vinther},
  {Ballester}, \& {Taylor}}]{Smette15}
{Smette}, A., {Sana}, H., {Noll}, S., {et~al.} 2015, \aap, 576, A77

\bibitem[{{Sollerman} {et~al.}(2005){Sollerman}, {Cox}, {Mattila},
  {Ehrenfreund}, {Kaper}, {Leibundgut}, \& {Lundqvist}}]{Sollerman05}
{Sollerman}, J., {Cox}, N., {Mattila}, S., {et~al.} 2005, \aap, 429, 559

\bibitem[{{Sousa} {et~al.}(2008){Sousa}, {Santos}, {Mayor}, {Udry},
  {Casagrande}, {Israelian}, {Pepe}, {Queloz}, \& {Monteiro}}]{Sousa08}
{Sousa}, S.~G., {Santos}, N.~C., {Mayor}, M., {et~al.} 2008, \aap, 487, 373

\bibitem[{{van Loon} {et~al.}(2013){van Loon}, {Bailey}, {Tatton}, {Ma{\'{\i}}z
  Apell{\'a}niz}, {Crowther}, {de Koter}, {Evans}, {H{\'e}nault-Brunet},
  {Howarth}, {Richter}, {Sana}, {Sim{\'o}n-D{\'{\i}}az}, {Taylor}, \&
  {Walborn}}]{vanLoon13}
{van Loon}, J.~T., {Bailey}, M., {Tatton}, B.~L., {et~al.} 2013, \aap, 550,
  A108

\bibitem[{{Vos} {et~al.}(2011){Vos}, {Cox}, {Kaper}, {Spaans}, \&
  {Ehrenfreund}}]{Vos11}
{Vos}, D.~A.~I., {Cox}, N.~L.~J., {Kaper}, L., {Spaans}, M., \& {Ehrenfreund},
  P. 2011, \aap, 533, A129

\bibitem[{{Welty} {et~al.}(2006){Welty}, {Federman}, {Gredel}, {Thorburn}, \&
  {Lambert}}]{Welty06}
{Welty}, D.~E., {Federman}, S.~R., {Gredel}, R., {Thorburn}, J.~A., \&
  {Lambert}, D.~L. 2006, \apjs, 165, 138

\bibitem[{{Xiang} {et~al.}(2012){Xiang}, {Liu}, \& {Yang}}]{xia12}
{Xiang}, F., {Liu}, Z., \& {Yang}, X. 2012, \pasj, 64, 31

\bibitem[{{Yuan} \& {Liu}(2012)}]{Yuan12}
{Yuan}, H.~B. \& {Liu}, X.~W. 2012, \mnras, 425, 1763

\bibitem[{{Zasowski} {et~al.}(2015){Zasowski}, {M{\'e}nard}, {Bizyaev},
  {Garc{\'{\i}}a-Hern{\'a}ndez}, {Garc{\'{\i}}a P{\'e}rez}, {Hayden},
  {Holtzman}, {Johnson}, {Kinemuchi}, {Majewski}, {Nidever}, {Shetrone}, \&
  {Wilson}}]{Zasowski15}
{Zasowski}, G., {M{\'e}nard}, B., {Bizyaev}, D., {et~al.} 2015, \apj, 798, 35

\end{thebibliography}

\end{document}